\documentclass{aa}
\usepackage{graphics}
\begin{document}

   \thesaurus{11     
              (11.01.2;  
               11.19.1;  
               13.09.1)} 

   \title{2.5--11 micron spectroscopy and imaging of AGNs:
   \thanks{Based on observations with ISO, an ESA projects with
           instruments funded by ESA member states (especially the
	   PI countries: France, Germany, the Netherlands and the United
	   Kingdom) and with the participation of ISAS and NASA} 
   }

   \subtitle{Implication for unification schemes}

   \author{J. Clavel \inst{1}
          \and
          B. Schulz\inst{2}
	  \and
	  B. Altieri \inst{1}
	  \and
	  P. Barr \inst{3}
	  \and
	  P. Claes \inst{3}
	  \and
	  A. Heras \inst{3}
	  \and
	  K. Leech \inst{2}
	  \and
	  L. Metcalfe \inst{2}
	  \and
	  A. Salama \inst{2}
          }

   \offprints{J. Clavel}

   \institute{XMM Science Operations, Astrophysics Division,
              ESA Space Science Dept., P.O. Box 50727, 28080 Madrid, Spain
              email: jclavel@xmm.vilspa.esa.es
          \and
	      ISO Data Centre, Astrophysics Division,
              ESA Space Science Dept., P.O. Box 50727, 28080 Madrid, Spain
          \and
	      Astrophysics Division, ESA Space Science Dept.,
	      ESTEC Postbus 299, 2200 AG -- Noordwijk, The Netherlands   
             }

   \date{ Received      ; accepted}

   \maketitle

   \begin{abstract}

   We present low resolution spectrophotometric and imaging ISO observations
   of a sample of 57 AGNs and one non-active SB galaxy over the 
   2.5--11~$\mu$m range. The sample is about equally divided into type I 
   ($\leq\,1.5$; 28 sources) and type II ($>\,1.5$; 29 sources) objects. 
   The mid-IR (MIR) spectra of type I (Sf1) and type II (Sf2) objects are 
   statistically {\em different\/}: Sf1 spectra are characterized by a 
   strong continuum well approximated by a power-law of average 
   index $\langle \alpha \rangle\,=\,-0.84\pm0.24$ with 
   only weak emission features from Polycyclic Aromatic Hydrocarbon (PAH) 
   bands at 3.3, 6.2, 7.7 and 8.6~$\mu$m. In sharp contrast to Sf1s, most 
   Sf2s display a weak continuum but very strong PAH emission bands, with
   equivalent widths (EW) up to 7.2~$\mu$m. On the other hand,
   Sf1s and Sf2s do not have statistically different PAH luminosities
   while the 7~$\mu$m continuum is on the average a factor $\sim$~8 less 
   luminous in Sf2s than in Sf1s. Because the PAH emission is unrelated 
   to the nuclear activity and arises in the interstellar medium of
   the underlying galactic bulge, its EW is a sensitive nuclear 
   redenning indicator. These results are consistent with unification 
   schemes and imply that the MIR nuclear continuum source of Sf2s is, 
   on the average, extinguished by $92\pm37$ visual magnitudes whereas 
   it is directly visible in Sf1s. The dispersion in Sf2's PAH EW is 
   consistent with the expected spread in viewing angles. Those Sf2s with 
   ${\rm EW(PAH)\,>\,5\,\mu m}$ suffer from an extinction ${\rm A_{v}\,>\,125}$ 
   magnitudes and are invariably extremely weak X-ray sources. Such Sf2s
   presumably represent the highly inclined objects where our line of
   sight intercepts the full extent of the molecular torus. Conversely,
   about a third of the Sf2s have PAH EW $\leq\,2\mu$m, in the range of
   Sf1s. Among them, those which have been observed in spectropolarimetry
   and/or in IR spectroscopy invariably display ``hidden'' broad lines. 
   As proposed by Heisler et al (\cite{heisler}), such Sf2s are most likely
   seen at grazing incidence such that one has a direct view of 
   {\em both\/} the ``reflecting screen'' {\em and\/} the torus inner 
   wall responsible for the near and mid-IR continuum. Our observations
   therefore constrain the screen and the torus inner wall to be
   spatially co-located. Finally, the 9.7~$\mu$m Silicate feature appears weakly
   in emission in Sf1s, implying that the torus vertical optical thickness
   cannot significantly exceed ${\rm 10^{24}\,cm^{-2}}$.
   \footnote{Tables 2,3,4 \& 5 are only available in electronic form at
   the CDS via anonymous ftp to cdsarc.u-strasbg.fr (130.79.128.5) or
   via http://cdsweb.u-strasbg.fr/Abstract.html.} 

   \end{abstract}

%

\section{Introduction}

   According to ``unified models'' of Active Galactic Nuclei (AGN), 
   Seyfert 1 and Seyfert 2 galaxies (hereafter Sf1 and Sf2) are essentially
   the same objects viewed at a different angle: Sf1s are observed 
   close to face-on such that we have a direct view to the Broad emission 
   Line Region (BLR) and the accretion disk responsible for the strong 
   UV-Optical-X-ray continuum, whereas Sf2s are seen at an inclination 
   such that our view is blocked by an optically thick dusty torus which 
   surrounds the disk and the BLR (e.g. Antonucci \cite{antonucci}). This 
   model makes specific predictions. In particular, the UV photons from the 
   disk which are absorbed by the grains in the torus should be re-emitted 
   as thermal radiation in the IR. Several arguments constrain the torus 
   inner radius to be of the order of $\sim$ 1 pc  in which case the 
   dust temperature should peak to about 700--1000~K and give rise to 
   an emission ``bump'' between $\simeq$ 2 and 15~$\mu$m (Pier \&
   Krolik \cite {pier}). The model also predicts that the silicate 
   9.7~$\mu$m feature should appear preferentially in absorption in Sf2s 
   and in emission in Sf1s. In order to test these predictions and better 
   constrain the model, we initiated a program of mid-IR (MIR) observations 
   of a large sample of AGNs. Throughout this paper, we use 
   ${\rm H_{0}\,=\,75\,km\,s^{-1};\,q_{0}\,=\,0}$. Unless otherwise
   stated, all quoted uncertainties correspond to 1-$\sigma$ errors.


\section{Observations}

  A sample of 57 AGNs and one non-active ``normal'' SB galaxy 
  were observed with the ISOPHOT (Lemke et al. \cite{lemke}) and ISOCAM 
  (Cesarsky et al. \cite{cesarskya}) instruments on board the Infrared 
  Space Observatory (ISO; Kessler et al. \cite{kessler}).
  
  Table~\ref{log} lists all the sources successfully observed with ISO.
  Columns 1--3 give the most common name and equatorial coordinates, columns
  4 \& 5 the seyfert type and the redshift, respectively, while columns 6--8
  list the instrument, the corresponding exposure time and start time of the
  observation. The redshifts and types are taken from the {\it NED\/}
  \footnote{The NASA/IPAC Extragalactic Database (NED) is operated 
  by the Jet Propulsion Laboratory, California Institute of Technology, 
  under contract with the National Aeronautics and Space Administration.}.
  The sample is drawn from the CfA hard X-ray flux limited complete sample
  (Piccinotti et al. \cite{piccinotti}) but lacks the most well known
  objects (e.g. \object{NGC~4151}) which were embargoed by ISO guaranteed 
  time owners. On the other hand, the sample was enriched in bright Sf2s.
  We caution that our sample is therefore not ``complete'' in a statistical 
  sense. It is about equally divided into Sf1s (28 sources, including 
  2 QSOs) and Sf2s (29), where we define Sf1s as all objects of type 
  $\leq$ 1.5 and Sf2s those whose type is $>$ 1.5. The mean and $r.m.s.$ 
  redshift are $0.047\pm0.083$ and $0.016\pm0.013$ for Sf1s and Sf2s. 
  Excluding the two QSO's (\object{HS~0624+6907} and \object{H~1821+643}), 
  the mean Sf1 redshift becomes 0.024$\pm$0.015, not significantly different
  from that of Sf2s. At these mean redshifts, a 10\,$\arcsec$ angular
  dimension projects onto a linear size at the source of 4.6 and 3.1 kpc, 
  for Sf1 and Sf2 respectively. 

  For every object but two, the data-set consists of ISOCAM images obtained   
  in staring mode through the LW2 and LW7 filters at 6.75 and 9.63~$\mu$m 
  respectively, with a 3~\arcsec per pixel magnification, together with 
  2.5--11.8~$\mu$m spectra obtained immediately before with the ISOPHOT-S 
  low resolution (${\rm 3360\,km\,s^{-1}}$) spectrograph. The images consist
  of arrays of $32\,\times\,32$ pixels (i.e. $96\,\times\,96$~\arcsec) with an
  effective resolution (FWHM) of 3.8~\arcsec and 4.5~\arcsec, for the LW2 and 
  LW7 filter, respectively. The exposure times per filter were always larger 
  than 200~s, sufficiently long to ensure proper stabilization of the 
  detectors. For the spectra, on-source measurements were alternated with 
  sky measurements at a frequency of 1/256~Hz, with a chopper throw of 
  $300~\arcsec$. Through a common $24~\arcsec\,\times\,24~\arcsec$ aperture, 
  light is fed simultaneously to the spectrograph and short wavelength 
  detector (ISOPHOT-SS: 2.5-4.8~$\mu$m) and to the equivalent long 
  wavelength channel (ISOPHOT-SL: 5.8--11.8~$\mu$m). There is a detector gap
  between 4.8 and 5.8~$\mu$m where no data can be recorded.
  \object{IR~05189-2524} was observed twice with the ISOCAM, while 
  \object{3C~390.3} and \object{Ark~564} were observed with ISOPHOT-S only. 
\section{Data reduction, calibration \& analysis}
\subsection{ISOCAM data}
  The ISOCAM images were reduced and calibrated using standard procedures of the
  CAM Interactive Analysis (CIA; Ott \cite{ott}) software package \footnote{CIA 
  is a joint development by the ESA Astrophysics Division and the ISOCAM 
  Consortium led by the ISOCAM PI, C. Cesarsky, Direction des Sciences 
  de la Matiere, C.E.A., France.} starting from the Edited Raw Data (ERD). 
  The first few ($\sim$ 4) readouts of each frame were discarded so as 
  to retain only those data for which the detector throughput had reached 
  90~\% of its stabilization level. The remainder of the processing involved 
  the usual steps, i.e. dark current subtraction, flat-fielding, removal 
  of particle hits (``de-glitching'') and flux calibration. All sources
  are detected at a very high significance level. Fluxes were 
  obtained by integrating all the emission in a circle of 3 pixels 
  radius ($9~\arcsec$) and subtraction of the background emission summed 
  over an external annulus centered on the source. These fluxes were 
  further multiplied by 1.23 to account for the emission in the wings
  of the PSF. In the case of extended sources (see below), the flux 
  was integrated over a circle of radius of 4.5 pixels (13.5\,$\arcsec$) 
  with the same area as the ISOPHOT-S aperture. The resulting ISOCAM fluxes 
  are listed in columns 2 and 4 of Table~\ref{cam_pht_photom}, where 
  an ``E'' in the last column denotes extended sources. The flux 
  accuracy is mainly limited by flat-fielding residuals and imperfect 
  stabilization of the signal. It is typically $\pm5$~\% for fluxes 
  greater than 200~mJy, 10~\% for fluxes in the range 100--200~mJy and 
  $\pm$15~\% at lower intensity levels. This is confirmed by the two 
  observations of the bright galaxy \object{IR~05189-2524} for which 
  the fluxes differ by only 5.5~\% and 0.5~\%, at 6.75~$\mu$m and 
  9.63~$\mu$m respectively  (see table~\ref{cam_pht_photom}). 

\subsection{ISOPHOT-S data}
  The ISOPHOT-S data were reduced with the PHOT Interactive Analysis (PIA;
  \footnote{PIA is a joint development by the ESA Astrophysics Division and 
  the ISOPHOT Consortium led by the ISOPHOT PI, D. Lemke, MPIA, Heidelberg.} 
  Gabriel \cite{gabriel}) software package. However, because  ISOPHOT-S was 
  operating close to its sensitivity limit, special reduction and 
  calibration procedures had to be applied. 
  During the measurement, the chopper mirror switches periodically between
  source and background. After such a change of illumination, 
  the photocurrent of the Si:Ga photoconductors immediately changes to an 
  intermediate level, followed by a slow characteristic 
  transition to the final level. At the fainter fluxes, a few Janskys and below,
  the time constant of this transition is extremely long. In our case 
  of chopped-mode observations with a frequency of $1/256$~Hz, 
  the final asymptotic value is never reached and only the initial steps towards the 
  final value are observed. These are practically equal to the intermediate level.
\begin{figure}
\resizebox{\hsize}{!}{\includegraphics{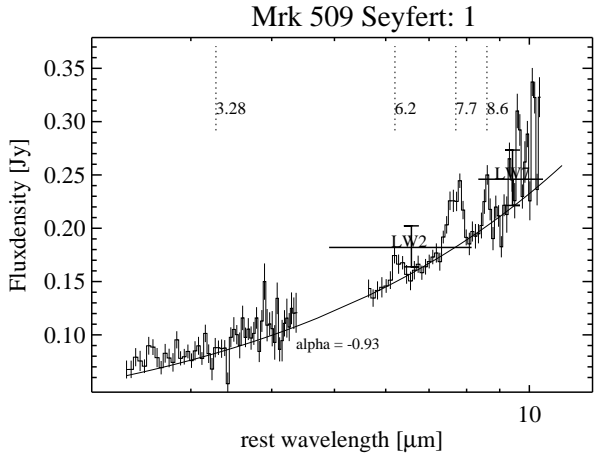}}
\caption{Representative spectrum of a Sf1 galaxy (Mrk~509) with the
  expected position of the PAH emission bands marked by vertical lines.
  Each flux data-point appears with its uncertainty.  
  The two large crosses marked LW2 and LW7 indicate the flux from the 
  ISOCAM images with its error and the filter wavelength range. The
  best-fit power-law continuum is shown as a continuous line.}
\label{mrk509}
\end{figure}
  This allows to simplify the flux calibration procedure
  by determining a spectral response function for this particular mode and 
  flux, thereby minimising possible systematic errors due to differences 
  in instrument configuration between the observations of calibrator and galaxy. 
  We derived this function from a chopped observation of a faint standard star 
  HD~132142 (TDT 63600901) of similar brightness to our objects, whose flux ranges 
  from 0.15 to 2.54 Jy. 
  The measurement was performed with the same chopper 
  frequency and readout-timing as the AGN observations.
  The $S/N$ of the ISOPHOT-S spectra was considerably enhanced by two 
  additional measures: 
  $i$)~the 32-s integration ramps were divided into sub-ramps of 2~sec,
  to provide enough signals per chopper plateau for the statistical analysis 
  and to keep the loss of integration time low when de-glitching,
  i.e. when removing ramps affected by cosmic ray hits.
  $ii$)~after fitting slopes to all sub-ramps, and removal
  of outliers (de-glitching) with PIA, the mean signal (slope) for each pixel 
  was determined separately for on- and off-source pointings by fitting 
  gaussians to the signal histograms.
  This corrects for the remaining asymmetry in the signal distribution, 
  i.e. the tail of the distribution towards higher values due to 
  non-recognised glitches is ignored and the result is closer to the median.
  We did not use the median itself because the digitisation of the voltages
  leads to signal-quantisation effects at small fluxes. 
  The difference of on- and off-source signals was then divided by our
  spectral response function to derive fluxes expressed in Janskys.
  Taking into account the accuracy of the model-SED of the stellar calibrator
  (Hammersley 1995, priv. comm., see also Hammersley \cite{hamm98}) 
  and the reproducibility of the measurements (Schulz \cite{schulz99}), the 
  flux calibration is accurate to within $\pm10$~\% (1-$\sigma$), 
  except at wavelengths $>$10~$\mu$m, where the flux of HR~132142 is weakest 
  and the noise in the calibration measurement dominates. This agrees
  also with the accuracies given in Klaas et al. (\cite{klaas}).
  The individual spectra will be published in a separate paper (Schulz
  et al. \cite{schulz}) and can be provided on request.
\begin{figure}
\resizebox{\hsize}{!}{\includegraphics{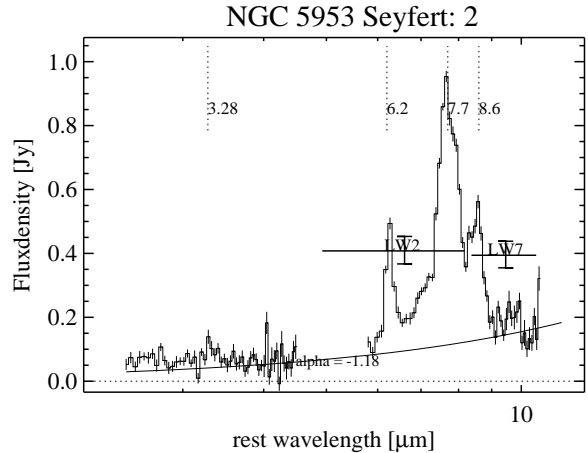}}
\caption{Same as Fig.~\ref{mrk509} but for the typical Sf2 galaxy NGC~5953.
Notice the much stronger PAH emission and weaker continuum}
\label{ngc5953}
\end{figure}
\subsection{Comparison of ISOCAM \& ISOPHOT-S fluxes}
  To check the reliability of our calibration, the ISOCAM and ISOPHOT-S
  results were compared as follows. The total ISOPHOT-S flux in the LW2
  and LW7 bands were computed by integration of the signal over the nominal 
  band-pass of the filters, 5.00--8.50~$\mu$m and 8.50--10.70~$\mu$m, 
  respectively. The exact shape of the filter spectral response is somewhat 
  uncertain and the LW2 band-pass extends over the data-gap between ISOPHOT-SS 
  and ISOPHOT-SL detector arrays. The integration was performed assuming
  a simple rectangular profile for the filter and interpolating the
  power-law continuum (see next section) over the 4.8--5.8~$\mu$m data gap.
  The resulting ISOPHOT-S fluxes are listed in columns 3 \& 5 of 
  Table~\ref{cam_pht_photom}. The projected linear dimension
  in kilo-parsec sustained by a 10\,$\arcsec$ angle is given
  for convenience in column 6. The ISOCAM and ISOPHOT-S  fluxes 
  are in good agreement, especially given the crudeness of the integration
  method and the assumption about the data gap. The mean ratios of the 
  ISOPHOT-S to the ISOCAM nuclear flux are 0.87 ($\pm$ 0.15, r.m.s.) and 
  1.07 ($\pm$ 0.20) at 6.75 and 9.63~$\mu$m respectively. Neither ratio
  differs significantly from unity. Our crude 
  interpolation method, which neglects emission above the continuum, 
  is at least partly responsible for the fact that the  6.75~$\mu$m ratio 
  is smaller than one (at the 0.9-$\sigma$ level only).  
  Averaged over the whole sample, the mean relative difference  
  between the ISOCAM and ISOPHOT-S fluxes amount to 16~\% for both filters.
  Selecting only those sources with a flux larger than 100~mJy, the 
  difference decreases to 14~\%, as expected. These figures are quite 
  close to the quadratic sum of the uncertainties on the ISOCAM and 
  ISOPHOT-S fluxes, thereby confirming the reliability of our 
  calibration and error estimates.
\begin{figure}
\resizebox{\hsize}{!}{\includegraphics{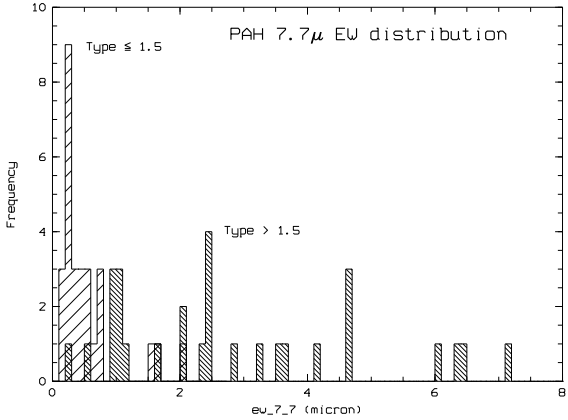}}
\caption{Distribution of the PAH 7.7~$\mu$m band EW 
for Sf1s (spaced hatching at 
+45~$\deg$) \& Sf2s (fine hatching, -45~$\deg$). 
}\label{hist_7pah_ew}
\end{figure}
\subsection{Extended sources}
  The 14 extended sources are listed in Table~\ref{ext_source} where
  columns 2--7 give the estimated nuclear flux, the flux in the ISOPHOT-S
  aperture and the total flux from the galaxy in the two ISOCAM 
  filters. Column 8 lists the spatial extension of the source obtained 
  by averaging the FWHM of the LW2 and LW7 images, while comments in 
  column 9 provide a brief morphological description of the object.
  In the case of \object{NGC~5953}, the galaxy is compact (4 pixels FWHM)
  with an approximately gaussian flux distribution and no well defined 
  point-source. In this case, no nuclear flux can be derived. For the
  other sources, the nuclear flux was obtained by deconvolution. Apart
  from \object{NGC~1097} which is clearly extended, all other sources are
  compact and smaller than the entrance of the ISOPHOT-S spectrograph.
  Averaged over the whole sample, the ratio of the nuclear flux to the
  flux in the ISOPHOT-S aperture is 0.77 and 0.79, at 6.75 and 9.63\,$\mu$m
  respectively. This implies that even for these extended sources, the 
  ISOPHOT-S spectrum is dominated by the nuclear emission, with a 
  20--25~\% contribution from the underlying galaxy. The mean ratio
  of the ISOPHOT-S flux to the total flux from the galaxy is 0.75 at
  6.75\,$\mu$m and 0.71 at 9.63\,$\mu$m. This further implies that the 
  extended MIR emission is relatively weak compared to the flux from 
  the central bulge and nucleus and that the bulk of it is anyway recorded 
  in the ISOPHOT-S spectrum. Hence, the redshift bias noted earlier should
  not have a significant impact on the ISOPHOT-S spectra.
\begin{figure}
\resizebox{\hsize}{!}{\includegraphics{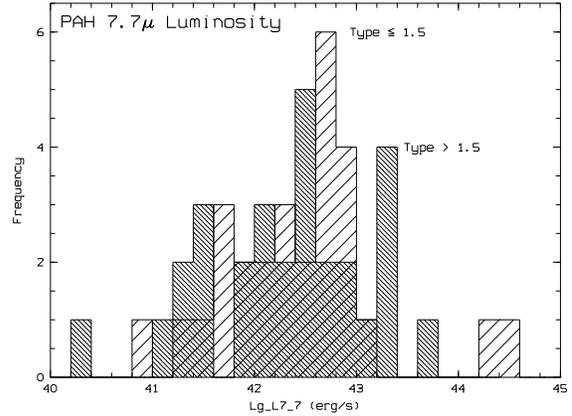}}
\caption{Distribution of the PAH 7.7~$\mu$m band luminosity
for Sf1s (spaced hatching at 
+45~$\deg$) \& Sf2s (fine hatching, -45~$\deg$). 
}\label{hist_7pah_lum}
\end{figure}
\subsection{Analysis}
  The typical spectrum of a Sf1 (\object{Mrk~509}) is shown in 
  Fig.~\ref{mrk509} and that of a Sf2 (\object{NGC~5953}) 
  in Fig.~\ref{ngc5953}. The MIR spectrum of a Sf2 is characterized by
  very strong emission features with  well defined peaks at 6.2, 7.7  
  and 8.6~$\mu$m, usually ascribed to Polycyclic Aromatic Hydrocarbon 
  (PAH) bands. The weaker 3.3~$\mu$m band is also detected in most sources
  and, in galaxies of adequate $S/N$ and redshift, the blue side
  of the strong 11.3~$\mu$m PAH feature also shows up as a sharp 
  rise in flux toward the long wavelength end of the ISOPHOT-S array.
  Though much weaker, the PAH emission bands are also present in Sf1s

  As can be judged from Fig.~\ref{mrk509}, the continuum of a Sf1 is
  well approximated by a power-law (${\rm F_{\nu}\,\propto\,\nu^{\alpha}}$).
  The continuum of a Sf2 is less well defined but in the absence of a
  better prescription and for the sake of consistency, we also adopted
  a power-law functional form to fit the continuum of type 2 sources. The
  results of the fit are given in Table~\ref{flux1}, were we list the spectral
  index and the continuum flux at a fiducial wavelength of 7~$\mu$m in 
  the rest frame of the object. No correction for foreground redenning
  in the milky way was applied as it is negligible at this wavelength.
  The error on the flux is estimated to be $\pm$10\,\% and the uncertainty
  on the spectral index $\pm$0.05 for Sf1s and $\pm$0.1 for Sf2s.
  
  The flux and equivalent widths (EW) of the PAH emission bands were measured
  by integrating all the flux above the best fit power-law continuum
  in the following pre-defined wavelength range (rest frame of the source):
  3.22--3.35\,$\mu$m, 5.86--6.54\,$\mu$m, 6.76--8.30\,$\mu$m and 
  8.30--8.90\,$\mu$m for the 3.3, 6.2, 7.7 and 8.6\,$\mu$m PAH band, 
  respectively. In addition, the total PAH flux (excluding the weakest
  3.3\,$\mu$m feature) was computed by integration over the range
  5.86--8.90\,$\mu$m. The results are given in Table~\ref{flux1} and 
  Table~\ref{flux2}. The errors quoted represent the quadratic sum of 
  the statistical uncertainties attached to each spectral data-point 
  in the integration interval. The ISOCAM images of \object{ESO~137-G34}
  show a star at a distance of 12\,$\arcsec$ from the nucleus, with a flux
  of 124 and 81~mJy at 6.75\,$\mu$m and 9.63\,$\mu$m respectively. Emission
  from this star therefore contaminates the ISOPHOT-S spectrum. This 
  contamination shows-up as an excess of short wavelength emission which 
  completely dominates the spectrum for $\lambda\,\leq\,4.5\,\mu$m.
  The parasitic spectrum was crudely estimated by fitting a straight
  line to the excess over the 2.5--4.5~$\mu$m range and removed from 
  the ISOPHOT-S spectrum of ESO~137-G34. For this object, the continuum 
  and line parameters are therefore subject to larger uncertainties than 
  the rest of the sample.

\section{The difference between Sf1 and Sf2}
  A comparison of Fig.~\ref{mrk509} and Fig.~\ref{ngc5953} reveals that the 
  MIR spectrum  of a typical Sf1 is markedly different from that of a Sf2:
  while Sf1s have a strong continuum  with only weak PAH emission, most 
  Sf2s are characterized by a weak continuum but very strong PAH emission.
  This difference is confirmed by a detailed statistical analysis. In Sf1s,
  the average equivalent width of the strongest of the PAH band at 7.7~$\mu$m
  is ${\rm \langle EW_{7.7} \rangle\,=\,0.53\pm0.47\,\mu m}$, where the error
  refers to the $r.m.s$ dispersion about the mean. This is 5.4 times
  smaller than the average equivalent width in Sf2s, 
  ${\rm \langle EW_{7.7} \rangle\,=\,2.86\pm1.95\,\mu m}$. Similarly, the
  equivalent width of the sum of the 3 strongest PAH features is
  ${\rm \langle EW_{TOT} \rangle\,=\,0.85\pm0.79\,\mu m}$ in Sf1s,
  compared to ${\rm \langle EW_{TOT} \rangle\,=\,4.38\pm2.98\,\mu m}$
  in Sf2s. The mean equivalent widths of the two populations and their
  variances are statistically different at the $10^{-7}$ and 
  $2\times 10^{-14}$ confidence level, respectively.
  
  The distribution of the $7.7~\mu$m PAH band EW is shown in 
  Fig.~\ref{hist_7pah_ew}. 
  It clearly illustrates that Sf1s and Sf2s have different EW 
  distributions: Sf1s are confined to a small range of EW with a maximum
  of 2.0~$\mu$m whereas Sf2s EW extend all the way up to a maximum of
  7.2~$\mu$m. A two-tail Kolmogorov-Smirnov (KS) test confirms that the
  Sf1 and Sf2 EW distributions are statistically different at the 
  $4\times 10^{-8}$ confidence level.  
  
  As can be seen from Fig.~\ref{hist_7pah_lum} however, the distribution 
  of the 7.7~$\mu$m PAH {\em luminosity\/} is the same for Sf1s and 
  Sf2s, at the 64~\% confidence level (KS test). The mean ($\pm r.m.s$) 
  7.7\,$\mu$m PAH luminosity of Sf1s is 
  ${\rm \langle \log{L_{7.7}} \rangle\,=\,42.44\pm\,0.80~erg\,s^{-1}}$, 
  not statistically different (at the 39~\% confidence level) from that of 
  Sf2s, ${\rm \langle \log{L_{7.7}} \rangle\,=\,42.28\pm0.78~erg\,s^{-1}}$.
  
  On the other hand, the average 7\,$\mu$m continuum luminosity of Sf1s
  ${\rm \langle \log{\nu\,L_{\nu,7}} \rangle=42.84\pm0.75}$ is
  nearly 8 times smaller than that of Sf2s, 
  ${\rm \langle \log{\nu\,L_{\nu,7}} \rangle=43.73\pm0.85}$.
  A KS test confirms that the luminosity distribution is also
  different for the two populations, at the $10^{-4}$ confidence
  level. The average continuum spectral index is however not statistically
  different in Sf1s (${\rm \langle \alpha \rangle = -0.84\pm0.24}$)
  and Sf2s (${\rm \langle \alpha \rangle = -0.82\pm0.37}$) and
  a KS test confirms that the distribution of indices is the same
  in the two populations at the 69~\% confidence level.
  
  It must be stressed that selection effects cannot account for the
  difference between Sf1s and Sf2s. First, the mean redshift and projected
  linear size of the two populations is not statistically different if 
  one excludes the two QSOs. Second, as noted in section 3.4, all sources
  are relatively compact and smaller than the entrance of the PHT-S 
  spectrograph, except NGC~1097. Third, the difference
  in PAH EW persists if one considers individual pairs of Sf1s and Sf2s with 
  similar redshift. For instance, the Sf1 \object{NGC~4051} and Sf2
  \object{NGC~5033} have very similar redshift (0.00242 versus 0.00292,
  respectively), but the 7.7\,$\mu$m PAH band EW of the latter, 2.394\,$\mu$m
  is 5.6 times larger than that of the former (0.428\,$\mu$m).
   
\section{Implications for unified schemes}   

  The simplest explanation of the above observational results is that 
  the MIR continuum is depressed in Sf2s relative to Sf1s. Indeed,
  a depressed continuum accounts for both the larger PAH EW and the
  reduced continuum luminosity of type 2 AGNs. This interpretation is 
  consistent with the weak anti-correlation that exists between the 
  7.7~$\mu$m PAH EW and the 7\,$\mu$m continuum luminosity, significant 
  at the $8\times 10^{-5}$ confidence level 
  (Kendall rank order coefficient = -0.358). The most likely reason 
  for the continuum depression is obscuration by dust, as postulated
  by unification schemes. The reddening law is essentially grey from 3
  to 11~$\mu$m and does not alter significantly the shape of the continuum.
  Hence, the obscuration hypothesis is consistent with Sf1 and Sf2 
  having the same spectral index on the average. On the other hand, 
  we have seen that the PAH luminosity
  is the same in Sf1s and Sf2s. This further requires that the region 
  responsible for the PAH emission is located outside the screen that 
  absorbs the MIR (and optical-UV) continuum. In other words, the
  screen must be located in the immediate vicinity of the nucleus such
  that it absorbs the 7~$\mu$m continuum but leaves the PAH emission
  unaffected. Again, this is consistent with unification schemes and
  with the molecular torus hypothesis. We have made here the implicit 
  assumption that the MIR continuum observed in Sf1s is of nuclear
  origin. There is strong observational support for this hypothesis:
\begin{itemize}  
  \item{Up to at least 3.4\,$\mu$m, the near-IR continuum of AGNs is 
  known to be variable, indicating a compact emission source 
  (e.g. Neugebauer et al. \cite{neugebauer}). Furthermore, in
  \object{F~9} (Clavel Wamsteker \& Glass \cite{clavel}), 
  \object{GQ~COMAE}  (Sitko, Sitko \& Siemiginowska \cite{sitko}), 
  \object{NGC~1566} (Baribaud et al. \cite{baribaud}) and 
  \object{NGC~3783} (Glass \cite{glass}), the variations follow
  closely those of the optical and ultraviolet continuum with a 
  delay of a few months to $\sim$ one year, which is commensurate 
  with the photon travel time to the dust sublimation radius. Such 
  observations are a strong indication that the 1.25--3.4~$\mu$m 
  flux of radio-quiet AGNs originates from thermal 
  emission by dust grains located in the immediate vicinity of the 
  central engine, presumably at the torus inner edge.}
  \item{AGNs have warm 12 to 100~$\mu$m colors, clearly different from those
  of ``normal'' non-active galaxies (e.g. Miley Neugebauer \& Soifer
  \cite{miley}) which indicates that the 12~$\mu$m emission is 
  associated with the nuclear activity}
\end{itemize}

  In the following, we have therefore assumed that the MIR continuum
  of Sf1s originates from thermal emission by dust grains located
  at the inner edge of the molecular torus.
  
  This leaves the origin of the PAH emission unclear. PAH emission
  is ubiquitous throughout the interstellar medium of our galaxy,
  in particular in star forming region and reflection nebulae
  (Verstraete et al. \cite{verstraete}; Cesarsky et al. 
  \cite{cesarskyb}; Cesarsky et al \cite{cesarskyc}) and galactic
  cirrus (Mattila et al. \cite{mattila}). PAH emission bands are 
  also conspicuous in normal late type galaxies (Xu et al. \cite{xu};
  Boulade et al. \cite{boulade}; Vigroux et al. \cite{vigroux}; 
  Metcalfe et al. \cite{metcalfe}; Acosta-Pulido et al. \cite{acosta}).
  Hence, the PAH emission bands seen in AGN spectra most probably arise
  in the general ISM of the underlying galaxies, more specifically in
  its bulge given that nearly all sources have low redshifts and are 
  unresolved at the 4--5~\arcsec resolution of ISOCAM. 
  
\begin{figure}
\resizebox{\hsize}{!}{\includegraphics{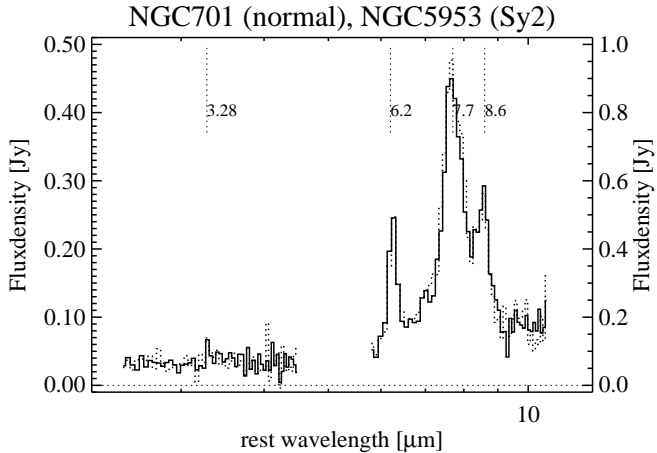}}
\caption{The spectra of the non-active SB galaxy NGC~701 
(heavy line) and of the Sf2 galaxy NGC~5953 (dotted-line)
The flux of NGC~701 has been divided by a factor 2.
}\label{n701_n5953}
\end{figure}

  To check this hypothesis, we overlay in Fig.~\ref{n701_n5953} the 
  spectrum of the typical Sf2 spectrum of \object{NGC~5953} on top of 
  the spectrum of the normal SB galaxy \object{NGC~701} (scaled by a 
  factor 0.5). The two spectra are virtually indistinguishable. Moreover,
  the 7.7~$\mu$m PAH EW in NGC~701, 5.88$\pm0.05\,\mu$m as well as the 
  luminosity, ${\rm \log{L_{7.7}}\,=\,42.048\pm0.009~erg\,s^{-1}}$, are 
  well within the range spanned by Sf2s. This confirms that
  the PAH emission is not related to the activity in the nucleus.
  Note also that the continuum in the two objects matches perfectly.
  This further implies that, at least in NGC~5953, the nuclear continuum
  is completely extinguished and the faint residual we observe originates
  from outside the active nucleus. The origin of this faint continuum
  in normal galaxies is currently a matter of debate. However, the fact 
  that it appears to correlate tightly with the PAH feature, possibly 
  indicate a common origin (Xu et al. \cite{xu}). Our results are at odds 
  with those of Malkan, Gorjian \& Tam (\cite{malkan}) who conclude that 
  Sf1s and Sf2s differ in galaxy type and infer that much of the extinction 
  occurs in the ISM, at larger radii than the molecular torus. We note
  however, that the statistical evidence for a difference in spectral 
  types between the host of Sf1s and Sf2s in the Malkan et al. sample is 
  marginal. 
\begin{figure}
\resizebox{\hsize}{!}{\includegraphics{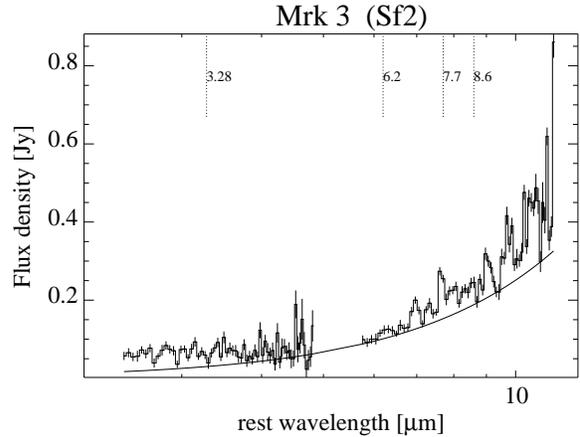}}
\caption{The spectrum of the Sf2 Mrk~3 with its best-fit power-law
continuum. Notice the weak Sf1-like PAH emission bands. 
}\label{mrk3}
\end{figure}
\section{The obscuration of Seyfert 2 galaxies}

  Since the PAH emission is not obscured by the torus while 
  the MIR continuum of Sf2s is, the PAH EW is a de-facto indicator
  of the nuclear extinction. According to unification schemes, the 
  extinction through which we see the nucleus depends on the orientation 
  of the torus with respect to our line of sight. One therefore expects 
  a spread in Sf2 PAH EW, as is indeed observed (Fig.~\ref{hist_7pah_ew}).
  The small EW sources correspond to the low inclination case where 
  our line of sight intercepts only the upper layers of the torus,
  while the large EW Sf2s are the sources seen close to edge-on 
  where our line of sight to the nucleus intercepts the full extent 
  of the torus. In such object, like NGC~5953, the MIR nuclear 
  emission is completely absorbed and we are left with the spectrum 
  of a normal galactic bulge, such as that of NGC~701. According to 
  unification schemes, the MIR continuum of Sf1s should not suffer  
  from extinction since we have a direct view of the torus inner edge. 
  We can therefore use the ratio R of the average PAH EW in Sf1s and
  in Sf2s to estimate the average MIR extinction of type 2 AGNs.
  This ratio for the strongest PAH band at 7.7~$\mu$m is
  $\langle R \rangle\,=\,5.4\pm3.7$, where the error quoted reflects  
  the $r.m.s.$ dispersion of Sf2 EWs. This implies that the continuum 
  of Sf2s suffers on the average from $1.83\pm0.74$ magnitudes of 
  extinction at 7.7~$\mu$m. This translates into a visual extinction
  ${\rm A_{v}\,=\,92\pm37}$ magnitudes 
  (Rieke and Lebofsky \cite{rieke}). For a normal gas to dust ratio, 
  this corresponds to an average X-ray absorbing column, 
  ${\rm N_{H}\,=\,2.0\pm0.8\times 10^{23}\,cm^{-2}}$ (Gorenstein \cite{gorenstein}). 
  The latter is in good agreement with the mean Sf2 absorbing column 
  as measured directly from X-ray data by Mulchaey et al. (\cite{mulchaey}), 
  ${\rm N_{H}\,=\,1.6^{+8.6}_{-1.3}\times 10^{23}\,cm^{-2}}$ or Smith 
  and Done (\cite{smith}), ${\rm N_{H}\,=\,1.0\pm1.3\times 10^{23}\,cm^{-2}}$.
  This excellent agreement should be seen more as a consistency check
  of our assumptions and a validation of unified schemes than an accurate
  determination of the torus optical depth. First, as mentioned earlier,
  it represents a mean value of the extinction averaged over the range
  of viewing angles, from grazing incidence to edge-on. Second, there is
  an intrinsic spread in the luminosity of the PAH emission, as illustrated
  by Fig.~\ref{hist_7pah_lum} which introduces uncertainties in this
  estimate. Third, in sources like NGC~5953 where the MIR continuum is
  totally absorbed, the PAH EW only provides a lower-limit of the true
  extinction.
  
\subsection{The highly obscured, large PAH EW Sf2s}
\begin{figure}
\resizebox{\hsize}{!}{\includegraphics{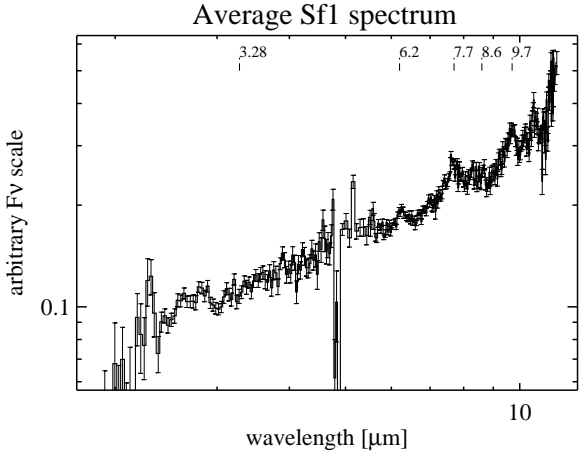}}
\caption{The average rest-frame spectrum of the 20 best signal-to-noise
 ratio Sf1 galaxies. The silicate 9.7~$\mu$m feature appears in emission. 
 }
\label{avg_sf1}
\end{figure}
  From the 29 Sf2s in our sample, 4 have ${\rm EW_{7.7}\,\geq\,5.5\,\mu m}$,
  in the range of a normal galaxy. This suggests that about 14~\% of Sf2s
  suffer from extinction in excess of 125 visual magnitudes, sufficient to 
  block-out the mid-IR continuum. These extreme Sf2s are presumably those 
  where the torus symmetry axis lies in the plane of the sky. The 4 galaxies
  are \object{IC~4397}, \object{ESO~137-G34}, \object{NGC~5728} and NGC~5953. 
  If our conclusions are correct, such extreme Sf2s should be heavily 
  absorbed in the X-rays, with hydrogen column densities in excess of  
  ${\rm 3\times 10^{23}\,cm^{-2}}$. Since for such cases the PAH EW ratio only 
  provides a lower limit to the extinction, these galaxies could even be 
  compton thick (i.e. ${\rm N_{H}\,\geq 10^{24}\,cm^{-2}}$) and opaque to 
  X-rays below $\sim$10~keV. In order to verify this prediction, we have 
  searched the literature and the NED for X-ray data on these sources. 
  IC~4397 and ESO~137-G34 have no entry in the ROSAT data-base nor in the 
  {\it EINSTEIN\/} catalog of IPC sources (Burstein et al. \cite{burstein}),
  presumably because they are too faint to have been detected. NGC~5728 is 
  not in the ROSAT data-base but appears in the IPC catalog with a 0.5--4.5~keV
  luminosity $\log{L_{x}}\,=\,40.92$, in the range of luminous non-active 
  galaxies. NGC~5953 coincides with the ROSAT HRI source 1RXH~J153432.5+151137.
  Its 0.5--2~keV luminosity $\log{L_{x}}\,=\,39.4$ can be accounted for 
  entirely by integrated stellar emission. It therefore appears that our 
  predictions are borne by existing X-ray data and that these 4 large 
  PAH EW Sf2s are indeed X-ray faint and probably compton thick. The 
  next three largest PAH EW Sf2s are \object{NGC~1667}, \object{Mrk~673} 
  and \object{NGC~5674}. They have 7.7~$\mu$m PAH EW of the order of 
  4.6~$\mu$m, smaller than that of the normal galaxy NGC~701, but still 
  a factor of $\sim$~10 larger than Sf1's. NGC~1667 was observed with 
  ASCA by Turner et al. (\cite{turnera}) who report luminosities of 
  ${\rm 0.8\times 10^{40}\,erg\,s^{-1}}$ and ${\rm 2.6\times 10^{40}\,erg\,s^{-1}}$ 
  over the 2--10~keV and 0.5--2~keV range, respectively, well within 
  the range of inactive galaxies. Furthermore, according to these authors,
\begin{figure}
\resizebox{\hsize}{!}{\includegraphics{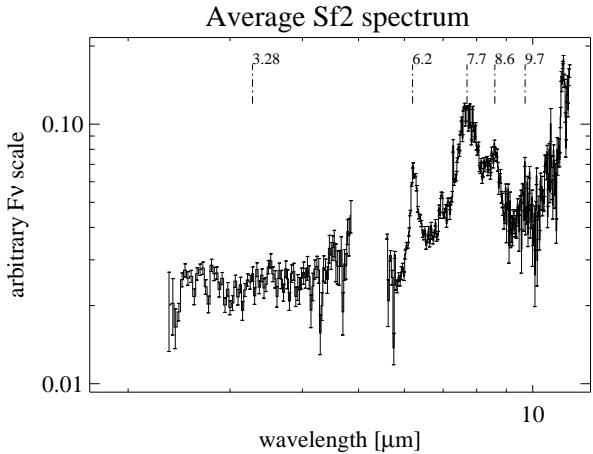}}
\caption{The average rest-frame spectrum of the 23 best signal-to-noise
 ratio Sf2 galaxies. 
 }
\label{avg_sf2}
\end{figure}
  68~\% of the X-ray flux in this galaxy can be accounted for by a thermal
  component due to the integrated stellar emission from a starburst.
  Mrk~673 appears in Burstein et al. (\cite{burstein}) with an upper
  limit ${\rm \log{L_{x}}\,\leq\,41.2}$ to its 0.5--4.5~keV luminosity. 
  Mrk~673 also coincides with the ROSAT PSPC source WGA~J1417.3+2651 with 
  a 0.5--2~keV luminosity ${\rm \log{L_{x}} = 40.8}$, again consistent 
  with a mostly stellar origin. Finally, NGC~5674 may represent the 
  first Sf2 where one starts seeing through the torus, since its X-ray 
  absorbing column could be measured with Ginga, 
  ${\rm 0.8\times 10^{23}\,cm^{-2}}$ (Smith \& Done \cite{smith}). To summarize,
  all large PAH EW Sf2s are faint and heavily absorbed in the X-rays,
  as predicted by unification schemes.

\subsection{The low obscuration, small PAH EW Sf2s}
  Ten Sf2 galaxies have 7.7~$\mu$m PAH EW $\leq\,2.0~\mu$m, in the range 
  occupied by Sf1s (fig~\ref{hist_7pah_ew}). The spectrum of Mrk~3 is 
  shown in Fig.\ref{mrk3} as an example of such a small PAH EW Sf2. 
  Among these ten Sf2s, four have been observed in spectropolarimetry 
  and/or near IR spectroscopy (Heisler et al. \cite{heisler}; Young et
  al. \cite{young}; Veilleux Goodrich \& Hill \cite{veilleux}). All 
  four galaxies display broad-lines in polarized light. These are 
  \object{Mrk~3}, \object{NGC~7674}, \object{IRAS~05189-2524} and 
  \object{NGC~4388}. Conversely, none of the three Sf2s with 7.7~$\mu$m 
  PAH EW $>\,1.6~\mu$m for which spectropolarimetric or IR spectroscopic 
  data exist (\object{Mrk~266}, \object{NGC~5728}, \object{NGC~1097}) 
  exhibit ``hidden'' broad lines. This confirms the finding of Heisler 
  et al. (\cite{heisler}) that the Sf2s which have ``hidden'' BLR 
  (i.e. seen in spectropolarimetry or in direct IR spectroscopy) are 
  those where our line-of-sight grazes the torus upper layer such that 
  we have a direct view of the reflecting mirror but not of the BLR. 
  As discussed previously, the MIR continuum most likely originates 
  from thermal emission by hot dust grains located on the inner wall 
  of the torus. Hence, the fact that the ``hidden'' BLR Sf2s are the 
  same sources which exhibit a Sf1-like MIR continuum further constrains 
  the mirror and the torus inner wall to be in neighbouring regions. It is 
  in fact conceivable that the mirror is the wall itself or a wind of hot
  electrons boiled-off the torus surface. It is interesting to note that 
  these ``hidden'' BLR Sf2s remain heavily absorbed in the X-rays while 
  their MIR continuum apparently does not suffer from a significant 
  amount of extinction. For instance, Mrk~3 has an X-ray absorbing column 
  ${\rm N_{H}\,=\,10^{24}\,cm^{-2}}$ (Turner et al \cite{turnerb}) 
  corresponding to a 7~$\mu$m extinction of 9 magnitudes, amply sufficient 
  to block-out the MIR continuum. Nevertheless, the MIR continuum of
  Mrk~3 (see Fig.\ref{mrk3}) is hardly absorbed. This indicates that our 
  line-of-sight to the X-ray source is different from our line-of-sight
  to the MIR source, the former intercepting a much larger fraction of
  the torus than the latter. As predicted by unified schemes, this 
  confirms that the X-ray source (like the BLR and the disk)
  is embedded further down the throat of the torus than the mirror 
  and the wall emitting the MIR continuum. 
  
  As a final remark, we note
  that MIR spectral characteristics do not allow to distinguish between 
  type 1.8, 1.9 and type 2 objects. For instance, the sub-class of small 
  PAH EW Sf2s include sources such as \object{Mrk~3} and \object{NGC~4507}
  which are bona-fide type 2 Seyferts. Conversely, \object{NGC~5674} and
  \object{Mrk~334} are classified as type 1.9 and 1.8 Seyfert galaxies 
  whereas their 7.7~$\mu$m PAH band EW is as large as 4.6~$\mu$m and 
  2.6~$\mu$m, respectively. Similarly, we do not find significant 
  differences between Broad-Line Radio-Galaxies (BLRG) and genuine
  radio-quiet Sf1s. Indeed, \object{3C~382} and \object{3C~390.3}
  have 7.7~$\mu$m PAH EW of 0.30 and 0.22\,$\mu$m respectively, close to
  the Sf1's average ($0.53\pm0.47\,\mu$m). Last, two sources in our sample
  qualify as ``Narrow-Line Seyfert 1'' (NLS1) galaxies, \object{Ark~564} 
  and \object{Mrk~507} (Boller, Brandt \& Fink \cite{boller}). With a
  7.7\,$\mu$m PAH EW of 0.26 and 0.59\,$\mu$m respectively, their MIR
  spectrum is undistinguishable from that of ``normal'' Sf1s. 
  
  Our scheme seems to be supported by the few published ISO spectra
  of Sf2 galaxies. The Circinus galaxy displays broad ${\rm H_{\alpha}}$
  emission (Oliva et al. \cite{oliva}) in polarized light while its MIR 
  spectrum reveals a strong continuum with relatively weak 
  (${\rm EW\,\leq\,2\,\mu m}$) PAH emission bands (Moorwood et al. 
  \cite{moorwood}). NGC~1068, the prototype Sf2 galaxy with a ``hidden BLR''
  also has weak PAH emission features (Genzel \cite{genzel}). Finally,
  IRAS~05189-2524 (Watson et al. \cite{watson}) has moderate strength 
  PAH emission bands (EW~$\simeq\,1\,\mu$m) together with polarized 
  broad-lines (Young et al. \cite{young}).
\section{The silicate 9.7 micron feature} 
  In individual spectra, there are hints of a spectral curvature around 
  9.7~$\mu$m, the expected wavelength of the silicate dust feature.
  However, the local continuum in the spectral range 
  $\lambda\,\geq\,9\,\mu$m is difficult to define due to the proximity
  of the 8.6~$\mu$m and 11.3~$\mu$m PAH emission bands. Moreover, the
  $S/N$ ratio decreases rapidly as one approaches the long wavelength
  end of the ISOPHOT-S array. Finally, the silicate feature itself appears
  to have a complex profile with a narrow absorption (possibly of 
  galactic origin) superimposed on a broader emission line. Attempts
  measuring its strength usually yield insignificant detections.
  To overcome this limit and increase the signal-to-noise ratio, we have
  computed the average MIR spectrum of a Sf1 and a Sf2 galaxy separately.
  Fig.~\ref{avg_sf1} shows the mean Sf1 spectrum obtained by normalizing
  and averaging the rest wavelength spectra of all 20 type $\leq\,1.5$ 
  AGNs with a signal-to-noise ratio per pixel large than 7, while 
  Fig.~\ref{avg_sf2} displays the average of the 23 Sf2 spectra with
  $\frac{S}{N}\,\geq\,3$. In the mean Sf1 spectrum, the 
  silicate 9.7~$\mu$m feature appears {\em in emission\/} with an 
  equivalent width ${\rm \langle EW_{9.7} \rangle\,=\,0.25\pm0.01\,\mu m}$. 
  This immediately {\em rules out models with very large torus optical 
  depths.\/} In the model of Pier and Krolik (\cite{pier}) for instance, 
  the strength of the silicate feature is calculated as a function of 
  inclination $i$ and of the vertical and radial Thomson optical depth,
  $\tau_{z}$ and $\tau_{r}$ respectively. Reading from  their figure 8, 
  models with $\tau_{z}\,\geq\,1$ and/or $\tau_{r}\,\geq\,1$
  are ruled-out as they predict the silicate feature in absorption.
  For an average Sf1 inclination $\cos{i}\,=\,0.8$, the best fit to 
  ${\rm \langle EW_{9.7} \rangle\,=\,0.254\pm0.008\,\mu m}$ suggests
  $\tau_{r}\,\simeq\,1$ and $0.1 \leq \tau_{z}\,\leq\,1$. A unit Thomson 
  optical depth corresponds to a column density 
  ${\rm N_{H}\,\simeq\,10^{24}\,cm^{-2}}$. While these figures are somewhat 
  model dependent, it is reassuring that they agree with our independent 
  estimate of ${\rm N_{H}}$ based on the PAH EW ratio.
  The PAH bands are so strong in Sf2s that placing 
  the continuum at 9.7~$\mu$m becomes a subjective decision. The mean Sf2 
  spectrum (Fig.~\ref{avg_sf2}) shows a weak maximum at 9.7~$\mu$m and a 
  shallow minimum near 10~$\mu$m. In the absence of longer wavelengths data, 
  one can only set a provisional upper limit of 0.32~$\mu$m to the silicate 
  EW in Sf2s, whether in absorption or in emission. It must be emphasized
  that the limitation is not the {\em S/N\/} but the uncertainty in 
  placing the local continuum.
\section{Summary and conclusions}

  A sample of 57 AGNs and one normal SB galaxy (NGC~701) were observed
  with the ISOPHOT-S spectrometer and the ISOCAM imaging camera. The
  sample is about equally divided into Sf1s (28) and Sf2s (29), where
  we define Sf1s as all objects of type $\leq 1.5$ and Sf2s those whose
  type is $>1.5$. The observations show that:
\begin{enumerate}
\item{Forty-four of the 57 AGNs in the sample appear unresolved at the
  $\simeq 4-5\arcsec$ resolution of ISOCAM. Of the 13 resolved sources,
  12 are sufficiently compact to ensure that all of the flux
  falls into the $24\arcsec\times\,24\arcsec$ ISOPHOT-S spectrograph 
  aperture. Moreover, even in these resolved sources, nuclear/bulge 
  emission contributes for at least 3/4\,th of the light recorded
  with ISOPHOT-S.}
\item{The spectrum of Sf1s is characterized by a strong continuum and
  weak Polycyclic Aromatic Hydrocarbon (PAH) emission bands at 3.3, 6.2, 
  7.7 and 8.6~$\mu$m. The continuum is well described by a power-law of 
  average index $\langle \alpha \rangle\,=-0.84\pm0.24$.}
\item{In sharp contrast with Sf1s, Sf2s generally have
  a weak continuum with very strong PAH emission bands.}
\item{The distribution of PAH equivalent widths (EW) is statistically 
  different in Sf1s and Sf2s. The average EW for the strongest band
  at 7.7~$\mu$m is $0.53\pm0.47~\mu$m in Sf1s versus $2.86\pm1.95~\mu$m
  in Sf2s. Moreover, the distribution of PAH EW in Sf1s is confined
  to values smaller than 2.0~$\mu$m whereas that of Sf2s extends from
  0.24~$\mu$m up to 7.2~$\mu$m.}
\item{There are however no statistical differences in the PAH luminosity
  distribution of Sf1s and Sf2s.}
\item{The 7~$\mu$m continuum is on the average a factor $\simeq$~8 less
  luminous in Sf2s than in Sf1s.}
\item{The PAH emission is not related to the activity in the nucleus and
  originates in the interstellar medium of the underlying galactic bulge.
  The PAH EW can therefore be used as a nuclear redenning indicator.}
\item{The above results are consistent with unification schemes and
  imply that the MIR continuum of Sf2s suffers from an average extinction
  of $92\pm37$ visual magnitudes. This corresponds to an average hydrogen
  absorbing column ${\rm N_{H}\,=\,2.0\pm0.8\,cm^{-2}}$, in good agreement
  with X-ray measurements. The large dispersion in the Sf2s EW is consistent
  with the expected spread in viewing angles.}
\item{The spectrum of Sf2s whose 7.7~$\mu$m PAH band EW exceeds 5~$\mu$m
  is indistinguishable from that of a normal non-active galaxy, implying
  that the MIR continuum is completely obscured in these sources 
  (${\rm A_{v}\,>\,125}$ magnitudes). Without exception, these Sf2s are 
  also heavily absorbed in the X-rays and probably ``compton thick''. These
  large PAH EW Sf2s are presumably those where the torus is seen 
  edge-on.}
\item{Ten Sf2s have 7.7~$\mu$m PAH EW $\leq\,2.0~\mu$m, in the range of
  Sf1s. Of these ten, four have been observed in spectropolarimetry
  and all four display ``hidden'' broad lines. Conversely, none of the
  three Sf2s with PAH EW $>\,2\,\mu$m which have been observed in
  spectropolarimtery display ``hidden'' broad lines. This confirms the 
  finding of Heisler et al. (\cite{heisler}) that those Sf2s with a 
  ``hidden'' BLR are those for which our line-of-sight grazes the upper
  surface of the torus. In these sources, we have a direct view of 
  {\em both\/} the reflecting mirror {\em and\/} of the torus inner 
  wall responsible for the MIR continuum. Thus, our observations strongly
  favour a model where the ``mirror'' and the torus inner wall are
  spatially co-located. It is in fact conceivable that the mirror is
  the torus inner wall itself or a wind of hot electrons boiled-off its
  surface by radiation pressure.}
\item{The silicate 9.7~$\mu$m feature appears weakly in emission in Sf1s.
  This implies that the torus cannot be extremely thick and the average
  silicate EW ($0.25\pm0.01\,\mu$m) suggests that the total hydrogen column 
  integrated along the torus vertical axis lies in the range 
  ${\rm 10^{23}\,\leq\,N_{H}\,\leq\,10^{24}\,cm^{-2}}$, consistent with
  our previous estimate based on the Sf2 PAH EW.}
\item{As far as their MIR properties are concerned, AGNs of intermediate 
  types 1.8 and 1.9 are indistingishable from genuine SF2s, whereas 
  Narrow Line Seyfert 1 (NLS1) and Broad-Line Radio-Galaxies (BLRG) 
  behave as normal SF1s.}
\end{enumerate}   

  The sketch outline in this paper makes specific predictions. First,
  Sf2s which have 7.7~$\mu$m PAH EW in excess of $\sim\,5\,\mu$m should
  never exhibit broad-lines in spectropolarimetry. Second, these sources
  should always be heavily absorbed in the X-rays, possibly up to 10~keV.
  Third, Sf2s whose PAH EW $\leq\,2\,\mu$m should exhibit broad lines when
  observed in spectropolarimetry and/or direct IR spectroscopy. This last 
  prediction seems to be borne by the few existing ISO observations of
  Sf2 with a ``hidden'' BLR


\begin{table*}
 \caption[]{Journal of ISO observations} 
  \begin{flushleft}
  \begin{tabular}{lcccccrc}
\hline
\hline
\\
TARGET    &        RA      &      Dec    & Type & z & INS & ${\rm T_{exp}}$ & Start Time \\
          &     (J2000)    &    (J2000)  &      &   &     &        (s)~     &   (UT)     \\
\\
\hline
\\
Mk 334       &  00h 03m 09.5s &+21d 57' 35.8" & 1.8 & 0.02196 & PHT & 2368 & 12 12 1996 00:22:43 \\
Mk 334       &  00h 03m 09.5s &+21d 57' 35.8" & 1.8 & 0.02196 & CAM &  568 & 12 12 1996 00:59:23 \\
Mk 335       &  00h 06m 19.5s &+20d 12' 10.7" & 1.0 & 0.02564 & PHT & 1344 & 12 12 1996 01:09:39 \\
Mk 335       &  00h 06m 19.5s &+20d 12' 10.7" & 1.0 & 0.02564 & CAM &  568 & 12 12 1996 01:29:13 \\
Fairall 9    &  01h 23m 45.9s &-58d 48' 21.0" & 1.0 & 0.04702 & PHT & 1344 & 06 05 1996 09:08:42 \\
Fairall 9    &  01h 23m 45.9s &-58d 48' 21.0" & 1.0 & 0.04702 & CAM &  568 & 06 05 1996 09:28:16 \\
NGC 526A     &  01h 23m 54.2s &-35d 03' 55.9" & 1.5 & 0.01922 & PHT & 2368 & 24 11 1996 02:37:19 \\
NGC 526A     &  01h 23m 54.2s &-35d 03' 55.9" & 1.5 & 0.01922 & CAM &  568 & 24 11 1996 03:13:59 \\
NGC 701      &  01h 51m 03.7s &-09d 42' 10.4" & SB  & 0.00610 & PHT & 4416 & 17 12 1997 23:24:55 \\
NGC 701      &  01h 51m 03.7s &-09d 42' 10.4" & SB  & 0.00610 & CAM &  568 & 18 12 1997 00:35:41 \\
Mk 590       &  02h 14m 33.6s &-00d 46' 00.0" & 1.5 & 0.02638 & PHT & 1344 & 03 07 1997 14:51:53 \\
Mk 590       &  02h 14m 33.6s &-00d 46' 00.0" & 1.5 & 0.02638 & CAM &  568 & 03 07 1997 15:11:27 \\
NGC 1097     &  02h 46m 19.1s &-30d 16' 28.2" & 2.0 & 0.00425 & PHT &  832 & 01 01 1997 02:09:46 \\
NGC 1097     &  02h 46m 19.1s &-30d 16' 28.2" & 2.0 & 0.00425 & CAM &  568 & 01 01 1997 02:20:48 \\
NGC 1125     &  02h 51m 40.4s &-16d 39' 02.2" & 2.0 & 0.01100 & PHT & 2368 & 01 02 1998 13:06:11 \\
NGC 1125     &  02h 51m 40.4s &-16d 39' 02.2" & 2.0 & 0.01100 & CAM &  568 & 01 02 1998 13:42:51 \\
NGC 1241     &  03h 11m 14.7s &-08d 55' 19.3" & 2.0 & 0.01351 & PHT & 4416 & 04 01 1998 13:52:47 \\
NGC 1241     &  03h 11m 14.7s &-08d 55' 19.3" & 2.0 & 0.01351 & CAM &  568 & 04 01 1998 15:03:35 \\
NGC 1386     &  03h 36m 46.2s &-35d 59' 58.0" & 2.0 & 0.00289 & PHT &  832 & 27 01 1998 02:39:25 \\
NGC 1386     &  03h 36m 46.2s &-35d 59' 58.0" & 2.0 & 0.00289 & CAM &  568 & 27 01 1998 02:50:27 \\
NGC 1566     &  04h 20m 00.7s &-54d 56' 17.3" & 1.0 & 0.00499 & PHT &  832 & 18 05 1997 17:57:48 \\
NGC 1566     &  04h 20m 00.7s &-54d 56' 17.3" & 1.0 & 0.00499 & CAM &  568 & 18 05 1997 18:08:52 \\
NGC 1667     &  04h 48m 36.9s &-06d 19' 12.9" & 2.0 & 0.01517 & PHT & 1344 & 02 10 1997 21:12:47 \\
NGC 1667     &  04h 48m 36.9s &-06d 19' 12.9" & 2.0 & 0.01517 & CAM &  568 & 02 10 1997 21:32:21 \\
Ark 120      &  05h 16m 11.4s &-00d 08' 59.4" & 1.0 & 0.03273 & PHT & 2368 & 20 08 1997 12:29:00 \\
Ark 120      &  05h 16m 11.4s &-00d 08' 59.4" & 1.0 & 0.03273 & CAM &  568 & 20 08 1997 13:05:38 \\
IR 05189-2524&  05h 21m 01.4s &-25d 21' 45.6" & 2.0 & 0.04256 & PHT & 2368 & 18 10 1997 00:20:57 \\
IR 05189-2524&  05h 21m 01.4s &-25d 21' 45.6" & 2.0 & 0.04256 & CAM &  568 & 18 10 1997 00:57:35 \\
IR 05189-2524&  05h 21m 01.4s &-25d 21' 45.6" & 2.0 & 0.04256 & CAM &  568 & 22 09 1997 08:04:39 \\
MCG 8-11-11  &  05h 54m 53.6s &+46d 26' 21.7" & 1.5 & 0.02048 & PHT & 2368 & 14 10 1997 23:15:13 \\
MCG 8-11-11  &  05h 54m 53.6s &+46d 26' 21.7" & 1.5 & 0.02048 & CAM &  568 & 14 10 1997 23:51:53 \\
Mrk 3        &  06h 15m 36.4s &+71d 02' 15.3" & 2.0 & 0.01351 & PHT & 2368 & 04 09 1997 10:35:21 \\
Mrk 3        &  06h 15m 36.4s &+71d 02' 15.3" & 2.0 & 0.01351 & CAM &  568 & 04 09 1997 11:12:01 \\
HS 0624+6907 &  06h 30m 02.7s &+69d 05' 04.8" & QSO & 0.37000 & PHT & 4416 & 04 09 1997 07:55:32 \\
HS 0624+6907 &  06h 30m 02.7s &+69d 05' 04.8" & QSO & 0.37000 & CAM &  568 & 04 09 1997 09:06:18 \\
NGC 3227     &  10h 23m 30.6s &+19d 51' 55.1" & 1.5 & 0.00386 & PHT &  832 & 25 04 1996 04:21:35 \\
NGC 3227     &  10h 23m 30.6s &+19d 51' 55.1" & 1.5 & 0.00386 & CAM &  568 & 25 04 1996 04:32:37 \\
A 1058+45    &  11h 01m 34.0s &+45d 39' 13.7" & 2.0 & 0.02908 & PHT & 4416 & 18 04 1996 08:40:29 \\
A 1058+45    &  11h 01m 34.0s &+45d 39' 13.7" & 2.0 & 0.02908 & CAM &  568 & 18 04 1996 09:51:17 \\
NGC 3516     &  11h 06m 47.5s &+72d 34' 06.6" & 1.5 & 0.00884 & PHT &  832 & 12 03 1996 11:21:20 \\
NGC 3516     &  11h 06m 47.5s &+72d 34' 06.6" & 1.5 & 0.00884 & CAM &  576 & 12 03 1996 11:32:22 \\
NGC 3982     &  11h 56m 28.2s &+55d 07' 29.3" & 2.0 & 0.00370 & PHT &  832 & 08 04 1996 06:03:32 \\
NGC 3982     &  11h 56m 28.4s &+55d 07' 29.3" & 2.0 & 0.00370 & CAM &  576 & 08 04 1996 06:14:36 \\
NGC 4051     &  12h 03m 09.6s &+44d 31' 52.9" & 1.0 & 0.00242 & PHT &  832 & 09 05 1996 09:25:53 \\
NGC 4051     &  12h 03m 09.6s &+44d 31' 52.9" & 1.0 & 0.00242 & CAM &  568 & 09 05 1996 09:36:57 \\
Mk 766       &  12h 18m 26.4s &+29d 48' 46.2" & 1.5 & 0.01293 & PHT &  832 & 02 06 1996 04:06:30 \\
Mk 766       &  12h 18m 26.4s &+29d 48' 46.2" & 1.5 & 0.01293 & CAM &  568 & 02 06 1996 04:17:32 \\
NGC 4388     &  12h 25m 46.7s &+12d 39' 40.7" & 2.0 & 0.00842 & PHT &  832 & 09 07 1996 06:07:32 \\
NGC 4388     &  12h 25m 46.7s &+12d 39' 40.7" & 2.0 & 0.00842 & CAM &  568 & 09 07 1996 06:18:36 \\
\\
\hline
\end{tabular}
\end{flushleft}
\label{log}
\end{table*}
\addtocounter{table}{-1}
\begin{table*}
 \caption[]{Journal of ISO observations (cont.)} 
  \begin{flushleft}
\begin{tabular}{lcccccrc}
\hline
\hline
\\
TARGET    &        RA      &      Dec    & Type & z & INS & ${\rm T_{exp}}$ & Start Time  \\
          &     (J2000)    &    (J2000)  &      &   &     &     (s)~        &   (UT)      \\
\\
\hline
\\
NGC 4507     &  12h 35m 36.7s &-39d 54' 33.4" & 2.0 & 0.01180 & PHT &  832 & 04 02 1996 11:55:40 \\
NGC 4507     &  12h 35m 36.7s &-39d 54' 33.4" & 2.0 & 0.01180 & CAM &  566 & 04 02 1996 12:06:42 \\
NGC 4579     &  12h 37m 43.5s &+11d 49' 04.6" & 1.9 & 0.00507 & PHT & 1344 & 12 07 1996 23:43:27 \\
NGC 4579     &  12h 37m 43.5s &+11d 49' 04.6" & 1.9 & 0.00507 & CAM &  568 & 13 07 1996 00:03:03 \\
NGC 4593     &  12h 39m 39.3s &-05d 20' 38.9" & 1.0 & 0.00900 & PHT & 1344 & 14 07 1996 06:00:28 \\
NGC 4593     &  12h 39m 39.3s &-05d 20' 38.9" & 1.0 & 0.00900 & CAM &  568 & 14 07 1996 06:20:04 \\
IR 12495-1308&  12h 52m 12.6s &-13d 24' 49.9" & 1.0 & 0.01463 & PHT & 2368 & 19 12 1996 10:52:42 \\
IR 12495-1308&  12h 52m 12.6s &-13d 24' 49.9" & 1.0 & 0.01463 & CAM &  568 & 19 12 1996 11:29:20 \\
NGC 5033     &  13h 13m 27.8s &+36d 35' 39.7" & 1.9 & 0.00292 & PHT &  832 & 25 06 1996 12:02:21 \\
NGC 5033     &  13h 13m 27.8s &+36d 35' 39.7" & 1.9 & 0.00292 & CAM &  568 & 25 06 1996 12:13:25 \\
MGC-6-30-15  &  13h 35m 53.7s &-34d 17' 45.3" & 1.0 & 0.00775 & PHT &  832 & 14 02 1996 17:40:28 \\
MGC-6-30-15  &  13h 35m 53.7s &-34d 17' 45.3" & 1.0 & 0.00775 & CAM &  566 & 14 02 1996 17:51:30 \\
Z 1335.5+3925&  13h 37m 39.9s &+39d 09' 16.4" & 1.8 & 0.02009 & PHT & 4416 & 09 06 1996 11:34:33 \\
Z 1335.5+3925&  13h 37m 39.9s &+39d 09' 16.4" & 1.8 & 0.02009 & CAM &  568 & 09 06 1996 12:45:19 \\
Mrk 266 SW   &  13h 38m 17.4s &+48d 16' 36.4" & 2.0 & 0.02786 & PHT & 1344 & 04 05 1996 15:44:50 \\
Mrk 266 SW   &  13h 38m 17.4s &+48d 16' 36.4" & 2.0 & 0.02786 & CAM &  568 & 04 05 1996 16:04:24 \\
NGC 5273     &  13h 42m 08.3s &+35d 39' 15.0" & 1.9 & 0.00352 & PHT & 2368 & 25 06 1996 11:14:43 \\
NGC 5273     &  13h 42m 08.3s &+35d 39' 15.0" & 1.9 & 0.00352 & CAM &  568 & 25 06 1996 11:51:23 \\
IC 4329 A    &  13h 49m 19.2s &-30d 18' 34.6" & 1.0 & 0.01605 & PHT &  832 & 14 02 1996 16:13:56 \\
IC 4329 A    &  13h 49m 19.2s &-30d 18' 34.6" & 1.0 & 0.01605 & CAM &  566 & 14 02 1996 16:25:00 \\
Mrk 279      &  13h 53m 03.4s &+69d 18' 29.1" & 1.0 & 0.02940 & PHT & 2368 & 05 02 1996 06:50:17 \\
Mrk 279      &  13h 53m 03.4s &+69d 18' 29.1" & 1.0 & 0.02940 & CAM &  566 & 05 02 1996 07:26:57 \\
Mrk 673      &  14h 17m 21.0s &+26d 51' 27.8" & 2.0 & 0.03651 & PHT & 2368 & 07 02 1996 19:18:22 \\
Mrk 673      &  14h 17m 21.0s &+26d 51' 27.8" & 2.0 & 0.03651 & CAM &  566 & 07 02 1996 19:55:02 \\
IC 4397      &  14h 17m 58.7s &+26d 24' 47.5" & 2.0 & 0.01474 & PHT & 4416 & 07 02 1996 17:57:42 \\
IC 4397      &  14h 17m 58.7s &+26d 24' 47.5" & 2.0 & 0.01474 & CAM &  566 & 07 02 1996 19:08:30 \\
NGC 5548     &  14h 17m 59.5s &+25d 08' 12.2" & 1.0 & 0.01717 & PHT & 1344 & 07 02 1996 16:37:46 \\
NGC 5548     &  14h 17m 59.5s &+25d 08' 12.2" & 1.0 & 0.01717 & CAM &  566 & 07 02 1996 16:57:22 \\
NGC 5674     &  14h 33m 52.1s &+05d 27' 30.7" & 1.9 & 0.02492 & PHT & 2368 & 07 02 1996 13:28:56 \\
NGC 5674     &  14h 33m 52.1s &+05d 27' 30.7" & 1.9 & 0.02492 & CAM &  566 & 07 02 1996 14:05:36 \\
Mk 817       &  14h 36m 22.0s &+58d 47' 39.0" & 1.5 & 0.03145 & PHT & 1344 & 05 02 1996 13:44:27 \\
Mk 817       &  14h 36m 22.0s &+58d 47' 39.0" & 1.5 & 0.03145 & CAM &  566 & 05 02 1996 14:04:02 \\
NGC 5728     &  14h 42m 23.8s &-17d 15' 09.9" & 2.0 & 0.00930 & PHT & 2368 & 07 02 1996 08:12:45 \\
NGC 5728     &  14h 42m 23.8s &-17d 15' 09.9" & 2.0 & 0.00930 & CAM &  566 & 07 02 1996 08:49:23 \\
Mk 841       &  15h 04m 01.1s &+10d 26' 16.0" & 1.0 & 0.03620 & PHT & 2368 & 07 02 1996 15:17:22 \\
Mk 841       &  15h 04m 01.1s &+10d 26' 16.0" & 1.0 & 0.03620 & CAM &  566 & 07 02 1996 15:54:02 \\
NGC 5929     &  15h 26m 06.1s &+41d 40' 14.5" & 2.0 & 0.00854 & PHT & 1344 & 05 02 1996 13:12:15 \\
NGC 5929     &  15h 26m 06.1s &+41d 40' 14.5" & 2.0 & 0.00854 & CAM &  566 & 05 02 1996 13:31:51 \\
NGC 5940     &  15h 31m 18.0s &+07d 27' 27.8" & 1.0 & 0.03405 & PHT & 2368 & 07 02 1996 14:29:36 \\
NGC 5940     &  15h 31m 18.0s &+07d 27' 27.8" & 1.0 & 0.03405 & CAM &  566 & 07 02 1996 15:06:16 \\
NGC 5953     &  15h 34m 32.4s &+15d 11' 37.8" & 2.0 & 0.00656 & PHT & 1344 & 07 02 1996 16:05:22 \\
NGC 5953     &  15h 34m 32.4s &+15d 11' 37.8" & 2.0 & 0.00656 & CAM &  566 & 07 02 1996 16:24:58 \\
ESO 137-G34  &  16h 35m 14.2s &-58d 04' 48.9" & 2.0 & 0.00916 & PHT & 2368 & 09 02 1996 07:54:40 \\
ESO 137-G34  &  16h 35m 14.2s &-58d 04' 48.9" & 2.0 & 0.00916 & CAM &  566 & 09 02 1996 08:31:18 \\
MRK 507      &  17h 48m 38.4s &+68d 42' 15.9" & 2.0 & 0.05590 & PHT & 4416 & 29 05 1997 14:30:14 \\
MRK 507      &  17h 48m 38.4s &+68d 42' 15.9" & 2.0 & 0.05590 & CAM & 1214 & 10 10 1997 21:20:24 \\
H 1821+643   &  18h 21m 57.2s &+64d 20' 36.4" & QSO & 0.29700 & PHT & 2368 & 05 02 1996 09:34:27 \\
H 1821+643   &  18h 21m 57.2s &+64d 20' 36.4" & QSO & 0.29700 & CAM &  566 & 05 02 1996 10:11:07 \\
\\
\hline
\end{tabular}
\end{flushleft}
\end{table*}
\addtocounter{table}{-1}
\begin{table*}
 \caption[]{Journal of ISO observations (cont.)} 
  \begin{flushleft}
\begin{tabular}{lcccccrc}
\hline
\hline
\\
TARGET    &        RA      &      Dec    & Type & z & INS & ${\rm T_{exp}}$ & Start Time  \\
          &     (J2000)    &    (J2000)  &      &   &     &     (s)~        &   (UT)      \\
\\
\hline
\\
3C 382       &  18h 35m 03.3s &+32d 41' 46.8" & 1.0 & 0.05787 & PHT & 4416 & 16 02 1996 17:19:38 \\
3C 382       &  18h 35m 03.3s &+32d 41' 46.8" & 1.0 & 0.05787 & CAM &  566 & 16 02 1996 18:30:26 \\
3C390.3      &  18h 42m 08.7s &+79d 46' 16.8" & 1.0 & 0.05610 & PHT & 4416 & 05 02 1996 04:38:11 \\
ESO 141-G55  &  19h 21m 14.2s &-58d 40' 12.4" & 1.0 & 0.03600 & PHT & 2368 & 04 03 1996 17:27:35 \\
ESO 141-G55  &  19h 21m 14.2s &-58d 40' 12.4" & 1.0 & 0.03600 & CAM &  576 & 04 03 1996 18:04:15 \\
Mk 509       &  20h 44m 09.7s &-10d 43' 24.5" & 1.0 & 0.03440 & PHT & 4416 & 18 10 1996 01:52:36 \\
Mk 509       &  20h 44m 09.7s &-10d 43' 24.5" & 1.0 & 0.03440 & CAM &  568 & 18 10 1996 03:03:22 \\
NGC 7314     &  22h 35m 45.7s &-26d 03' 02.9" & 1.9 & 0.00474 & PHT & 2368 & 29 04 1996 12:05:36 \\
NGC 7314     &  22h 35m 45.7s &-26d 03' 02.9" & 1.9 & 0.00474 & CAM &  568 & 29 04 1996 12:42:14 \\
IR 22377+0747&  22h 40m 17.6s &+08d 03' 15.4" & 1.8 & 0.02460 & PHT & 1344 & 18 05 1996 09:06:54 \\
IR 22377+0747&  22h 40m 17.6s &+08d 03' 15.4" & 1.8 & 0.02460 & CAM &  568 & 18 05 1996 09:26:28 \\
Ark 564      &  22h 42m 39.4s &+29d 43' 31.3" & 1.0 & 0.02400 & PHT & 4416 & 20 05 1997 14:28:34 \\
NGC 7592     &  23h 18m 22.1s &-04d 24' 59.7" & 2.0 & 0.02444 & PHT & 2368 & 15 11 1996 15:20:15 \\
NGC 7592     &  23h 18m 22.1s &-04d 24' 59.7" & 2.0 & 0.02444 & CAM &  568 & 15 11 1996 15:56:53 \\
NGC 7603     &  23h 18m 56.6s &+00d 14' 36.8" & 1.5 & 0.02952 & PHT & 1344 & 18 05 1996 05:27:37 \\
NGC 7603     &  23h 18m 56.6s &+00d 14' 36.8" & 1.5 & 0.02952 & CAM &  568 & 18 05 1996 05:47:11 \\
NGC 7674     &  23h 27m 56.7s &+08d 46' 43.5" & 2.0 & 0.02906 & PHT &  832 & 28 05 1996 06:28:21 \\
NGC 7674     &  23h 27m 56.7s &+08d 46' 43.5" & 2.0 & 0.02906 & CAM &  568 & 28 05 1996 06:39:23 \\
\\ 
\hline
\end{tabular}
\end{flushleft}
\end{table*}

\begin{table*}
\caption[]{6.75 \& 9.63\,$\mu$m CAM \& PHT fluxes}
\begin{tabular}{lcccccc}
\hline
\hline
\\
\multicolumn{1}{l}{TARGET} & 
\multicolumn{2}{c}{6.75\,$\mu$m nuclear flux} & 
\multicolumn{2}{c}{9.63\,$\mu$m nuclear flux} &
\multicolumn{1}{c}{${\rm l_{10}}$} &
\multicolumn{1}{c}{Note}\\
\multicolumn{1}{c}{}       &
\multicolumn{1}{c}{CAM}    & 
\multicolumn{1}{c}{PHT}    & 
\multicolumn{1}{c}{CAM}    & 
\multicolumn{1}{c}{PHT}    & 
\multicolumn{1}{c}{}       \\
\multicolumn{1}{c}{}       &
\multicolumn{1}{c}{(mJy)}  &
\multicolumn{1}{c}{(mJy)}  &
\multicolumn{1}{c}{(mJy)}  &
\multicolumn{1}{c}{(mJy)}  &
\multicolumn{1}{c}{(Kpc/10\,$\arcsec$)} &
\multicolumn{1}{c}{}       \\
\\
\hline
\\             
      Mrk 334 &    131 &    107 &    148 &  167 & 4.3 &   \\
      Mrk 335 &    155 &    153 &    203 &  225 & 4.3 &   \\
    Fairall 9 &    174 &    148 &    272 &  205 & 9.1 &   \\
     NGC 526a &    118 &    107 &    167 &  169 & 3.7 &   \\
     NGC 701  &    162 &    119 &    142 &  128 & 1.2 & E \\
     Mrk 590  &    144 &    123 &    198 &  215 & 5.1 &   \\
     NGC 1097 &   1212 &    729 &    825 &  813 & 0.8 & E \\
     NGC 1125 &     72 &     63 &     61 &   47 & 2.1 &   \\
     NGC 1241 &     67 &     43 &     43 &   44 & 2.6 & E \\
     NGC 1386 &    257 &    213 &    276 &  319 & 0.6 &   \\
     NGC 1566 &    116 &    157 &    114 &  190 & 1.0 & E \\
     NGC 1667 &    194 &    114 &    167 &  149 & 2.9 & E \\
      Ark 120 &    140 &    128 &    188 &  203 & 6.3 &   \\
IR 05189-2524 &    302 &     NA &    451 &   NA & 8.2 &   \\
IR 05189-2524 &    320 &    252 &    449 &  418 & 8.2 &   \\
  MCG 8-11-11 &    230 &    213 &    383 &  376 & 4.0 &   \\
        Mrk 3 &    154 &    117 &    320 &  302 & 2.6 &   \\
 HS 0624+6907 &     63 &     54 &     57 &   73 & 73. &   \\
     NGC 3227 &    294 &    249 &    382 &  372 & 0.7 &   \\
    A 1058+45 &     42 &     27 &     46 &   47 & 5.6 &   \\
     NGC 3516 &    264 &    263 &    369 &  392 & 1.7 &   \\
     NGC 3982 &    147 &    129 &    113 &  182 & 0.7 & E \\
     NGC 4051 &    265 &    262 &    411 &  454 & 0.5 &   \\
      Mrk 766 &    177 &    185 &    274 &  318 & 2.5 &   \\
     NGC 4388 &    265 &    256 &    267 &  361 & 1.6 &   \\
     NGC 4507 &    274 &    251 &    410 &  428 & 2.3 &   \\
     NGC 4579 &    104 &    114 &    113 &  159 & 1.0 &   \\
     NGC 4593 &    206 &    200 &    280 &  309 & 1.7 &   \\
IR 12495-1308 &     51 &     40 &     62 &   82 & 2.8 &   \\
     NGC 5033 &    239 &    200 &    210 &  225 & 0.6 & E \\
 MCG -6-30-15 &    213 &    206 &    329 &  388 & 1.5 &   \\
Z 1335.5+3925 &     49 &     39 &     49 &   59 & 3.9 & E \\
   Mrk 266 SW &    125 &     92 &    108 &  133 & 5.4 & E \\
     NGC 5273 &     23 &     25 &     25 &   37 & 0.7 &   \\
     IC 4329A &    591 &    477 &    890 &  739 & 3.1 &   \\
      Mrk 279 &    102 &     94 &    146 &  143 & 5.7 &   \\
      Mrk 673 &     72 &     55 &     73 &   80 & 7.1 &   \\
      IC 4397 &     64 &     44 &     59 &   64 & 2.9 & E \\
     NGC 5548 &    177 &    168 &    268 &  275 & 3.3 &   \\
     NGC 5674 &     38 &     26 &     43 &   50 & 4.8 & E \\
      Mrk 817 &    142 &    123 &    259 &  234 & 6.1 &   \\
     NGC 5728 &    147 &    114 &    126 &  114 & 1.8 & E \\
      Mrk 841 &     74 &     66 &    133 &  115 & 7.0 &   \\
     NGC 5929 &     41 &     30 &     72 &   51 & 1.7 &   \\
     NGC 5940 &     46 &     36 &     81 &   67 & 6.6 &   \\
     NGC 5953 &    310 &    247 &    243 &  252 & 1.3 & E \\
  ESO 137-G34 &     70 &     91 &     60 &   63 & 1.8 &   \\
      Mrk 507 &     29 &     19 &     52 &   36 & 11. &   \\
   H 1821+643 &    109 &     99 &    143 &  156 & 58. &   \\
\\
\hline
\end{tabular}
\label{cam_pht_photom}
\end{table*}
\addtocounter{table}{-1}
\begin{table*}
\caption[]{6.75 \& 9.63\,$\mu$m CAM \& PHT fluxes (cont.)}

\begin{tabular}{lcccccc}
\hline
\hline
\\
\multicolumn{1}{l}{TARGET} & 
\multicolumn{2}{c}{6.75\,$\mu$m nuclear flux} & 
\multicolumn{2}{c}{9.63\,$\mu$m nuclear flux} &
\multicolumn{1}{c}{${\rm l_{10}}$} &
\multicolumn{1}{c}{Note}\\
\multicolumn{1}{c}{}       &
\multicolumn{1}{c}{CAM}    & 
\multicolumn{1}{c}{PHT}    & 
\multicolumn{1}{c}{CAM}    & 
\multicolumn{1}{c}{PHT}    & 
\multicolumn{1}{c}{}       \\
\multicolumn{1}{c}{}       &
\multicolumn{1}{c}{(mJy)}  &
\multicolumn{1}{c}{(mJy)}  &
\multicolumn{1}{c}{(mJy)}  &
\multicolumn{1}{c}{(mJy)}  &
\multicolumn{1}{c}{(Kpc/10\,$\arcsec$)} &
\multicolumn{1}{c}{}       \\
\\
\hline
\\             
       3C 382 &     62 &     56 &     76 &	 72 & 11. &   \\
     3C 390.3 &     *  &     50 &      * &	 81 & 11. &   \\
  ESO 141-G55 &    103 &    100 &    145 &	184 & 7.0 &   \\
      Mrk 509 &    162 &    149 &    222 &	242 & 6.7 &   \\
     NGC 7314 &     51 &     51 &     67 &	 72 & 1.0 &   \\
IR 22377+0747 &     45 &     41 &     50 &	 64 & 4.8 &   \\
      Ark 564 &      - &     77 &      - &	132 & 4.7 &   \\
     NGC 7592 &    171 &    128 &    136 &	162 & 4.7 & E \\
     NGC 7603 &    138 &    109 &    176 &	152 & 5.7 &   \\
     NGC 7674 &    259 &    213 &    344 &      348 & 5.6 &   \\
\\
\hline
\end{tabular}
\end{table*}

\begin{table*}
\caption[]{Fluxes and dimensions of extended sources}
\begin{tabular}{lcccccccl}
\hline
\hline
\\
\multicolumn{1}{l}{TARGET}                & 
\multicolumn{3}{c}{6.75\,$\mu$m flux}      & 
\multicolumn{3}{c}{9.63\,$\mu$m flux}      &
\multicolumn{1}{c}{FWHM}         &
\multicolumn{1}{l}{Comments}              \\
\multicolumn{1}{c}{}                      &
\multicolumn{1}{c}{Nucl.}               & 
\multicolumn{1}{c}{$24\times24\,\arcsec$} & 
\multicolumn{1}{c}{Total}                 & 
\multicolumn{1}{c}{Nucl.}               & 
\multicolumn{1}{c}{$24\times24\,\arcsec$} &
\multicolumn{1}{c}{Total}                 &
\multicolumn{1}{l}{}                      &
\multicolumn{1}{l}{}                      \\
\multicolumn{1}{c}{}                      &
\multicolumn{1}{c}{(mJy)}                 &
\multicolumn{1}{c}{(mJy)}                 &
\multicolumn{1}{c}{(mJy)}                 &
\multicolumn{1}{c}{(mJy)}                 &
\multicolumn{1}{c}{(mJy)}                 &
\multicolumn{1}{c}{(mJy)}                 &
\multicolumn{1}{c}{$\arcsec$}             &
\multicolumn{1}{l}{}                      \\

\\
\hline
\\             
     NGC 701  &  119 &  162 &  258 &   92 & 142 &  250 & 8.0 & {\tiny Edge-on spiral; well defined bulge} \\
     NGC 1097 &  626 & 1212 & 1445 &  624 & 825 & 1024 &  28 & {\tiny Ill-defined nucleus; bright annular galaxy} \\
     NGC 1241 &   42 &   67 &  104 &   35 &  43 &  133 & 5.6 & {\tiny Nearly point-nucleus; faint annular galaxy} \\
     NGC 1566 &  109 &  116 &  210 &  112 & 114 &  190 & 4.7 & {\tiny Point-nucleus; faint spiral arms} \\
     NGC 1667 &   94 &  194 &  300 &   68 & 167 &  274 & 9.0 & {\tiny Faint bulge; bright oval galaxy} \\
     NGC 3982 &   62 &  147 &  262 &   68 & 113 &  300 & 4.5 & {\tiny Point-nucleus in circular galaxy} \\
     NGC 5033 &  179 &  239 &  600 &  150 & 210 &  675 & 8.5 & {\tiny Well defined nucleus; bright edge-on spiral} \\
Z 1335.5+3925 &   47 &   49 &   49 &   44 &  49 &   50 & 7.5 & {\tiny Compact oval galaxy; no defined nucleus} \\
   Mrk 266 SW &  114 &  125 &  130 &  102 & 108 &  108 & 4.5 & {\tiny Bright point-nucleus; faint compact galaxy} \\
      IC 4397 &   48 &   64 &   73 &   41 &  59 &   65 & 7.0 & {\tiny Faint point-nucleus; faint compact galaxy} \\
     NGC 5674 &   36 &   38 &   84 &   41 &  43 &   96 & 4.5 & {\tiny Point-nucleus; faint ring galaxy} \\
     NGC 5728 &  139 &  147 &  147 &  112 & 126 &  126 & 8.0 & {\tiny Nearly point-nucleus; compact galaxy} \\
     NGC 5953 &    - &  310 &  310 &    - & 243 &  243 &  12 & {\tiny No defined nucleus; compact galaxy} \\
     NGC 7592 &  171 &  171 &  171 &  136 & 136 &  136 & 4.5 & {\tiny Two point-nuclei separated by 9 arcsec} \\
\\
\hline
\end{tabular}
\label{ext_source}
\end{table*}

\begin{table*}
 \caption[]{Continuum and PAH band fluxes} 
  \begin{flushleft}
\begin{tabular}{lccccccc}
\hline
\hline
\\
TARGET        & Type&$\alpha$&F(7$\mu$m)&  F(3.3$\mu$m)  &EW(3.3$\mu$m)&  F(6.2$\mu$m)   &EW(6.2$\mu$m)\\
              &     &        & 
		(mJy) & 
	      ${\rm (10^{-12}\frac{erg}{cm^{2}\,s})}$ & 
		($\mu$m) & 
	      ${\rm (10^{-12}\frac{erg}{cm^{2}\,s})}$ & 
		($\mu$m) \\
\\
\hline
\\	       
Mrk 334       & 1.8 & -1.18  &  80.0 &     $\leq$0.260 &$\leq$0.03&  2.835$\pm$0.130 &  0.525\\
Mrk 335       & 1.0 & -0.50  & 152.6 & 0.450$\pm$0.390 &  0.030   &  1.082$\pm$0.180 &  0.086\\
Fairall 9     & 1.0 & -0.53  & 146.0 &     $\leq$0.299 &$\leq$0.01&  1.015$\pm$0.167 &  0.101\\
NGC 526a      & 1.5 & -0.72  &  94.7 & 0.609$\pm$0.258 &  0.041   &  2.247$\pm$0.117 &  0.322\\
NGC 701       &  SB & -0.84  &  48.9 & 0.633$\pm$0.154 &  0.099   &  4.312$\pm$0.078 &  1.299\\
Mrk 590       & 1.5 & -0.86  & 119.6 & 1.601$\pm$0.343 &  0.080   &  1.108$\pm$0.154 &  0.139\\
NGC 1097      & 2.0 & -0.96  & 402.8 & 4.202$\pm$0.456 &  0.082   & 20.158$\pm$0.321 &  0.744\\
NGC 1125      & 2.0 & -0.44  &  40.1 & 0.398$\pm$0.221 &  0.060   &  1.427$\pm$0.108 &  0.551\\
NGC 1241      & 2.0 & -0.95  &  32.7 & 1.391$\pm$0.159 &  0.314   &  0.639$\pm$0.077 &  0.295\\
NGC 1386      & 2.0 & -1.01  & 191.1 & 0.528$\pm$0.395 &  0.033   &  2.160$\pm$0.240 &  0.165\\
NGC 1566      & 1.0 & -0.85  & 112.9 & 6.099$\pm$0.668 &  0.372   &  3.697$\pm$0.494 &  0.423\\
NGC 1667      & 2.0 & -0.66  &  58.7 & 1.620$\pm$0.322 &  0.176   &  3.564$\pm$0.162 &  0.896\\
Ark 120       & 1.0 & -0.97  & 130.7 & 1.067$\pm$0.248 &  0.059   &  1.233$\pm$0.115 &  0.139\\
IR 05189-2524 & 2.0 & -0.92  & 229.3 & 0.703$\pm$0.215 &  0.024   &  3.350$\pm$0.110 &  0.215\\
MCG 8-11-11   & 1.5 & -0.81  & 223.2 & 0.597$\pm$0.230 &  0.017   &  0.678$\pm$0.119 &  0.037\\
Mrk 3         & 2.0 & -1.91  & 127.4 & 1.093$\pm$0.228 &  0.114   &  0.640$\pm$0.110 &  0.084\\
HS 0624+6907  & QSO & -0.77  &  65.2 & 0.492$\pm$0.225 &  0.078   &  0.370$\pm$0.122 &  0.072\\
NGC 3227      & 1.5 & -0.75  & 189.9 & 3.050$\pm$0.498 &  0.112   &  4.313$\pm$0.241 &  0.331\\
A 1058+45     & 2.0 & -0.93  &  16.8 & 0.310$\pm$0.173 &  0.162   &  0.707$\pm$0.077 &  0.650\\
NGC 3516      & 1.5 & -0.60  & 270.9 & 2.248$\pm$0.545 &  0.032   &  1.438$\pm$0.241 &  0.070\\
NGC 3982      & 2.0 & -0.65  &  86.1 & 2.175$\pm$0.440 &  0.161   &  2.547$\pm$0.245 &  0.418\\
NGC 4051      & 1.0 & -0.96  & 265.3 &     $\leq$0.427 &$\leq$0.04&  1.241$\pm$0.241 &  0.070\\
Mrk 766       & 1.5 & -0.90  & 198.5 & 1.111$\pm$0.508 &  0.005   &  0.477$\pm$0.222 &  0.032\\
NGC 4388      & 2.0 & -0.65  & 198.8 &     $\leq$0.649 &$\leq$0.04&  3.644$\pm$0.239 &  0.284\\
NGC 4507      & 2.0 & -1.15  & 268.5 & 0.804$\pm$0.441 &  0.049   &  1.853$\pm$0.224 &  0.105\\
NGC 4579      & 1.9 & -0.37  &  91.4 & 5.146$\pm$0.336 &  0.280   &  1.472$\pm$0.162 &  0.230\\
NGC 4593      & 1.0 & -0.52  & 199.2 & 0.664$\pm$0.393 &  0.023   &  0.818$\pm$0.163 &  0.053\\
IR 12495-1308 & 1.0 & -1.37  &  38.4 & 0.869$\pm$0.271 &  0.165   &  0.648$\pm$0.130 &  0.289\\
NGC 5033      & 1.9 & -0.75  & 132.8 & 3.577$\pm$0.438 &  0.198   &  4.517$\pm$0.220 &  0.497\\
MCG-6-30-15   & 1.0 & -0.84  & 218.4 & 0.590$\pm$0.406 &  0.037   &  0.083$\pm$0.247 &  0.012\\
Z 1335.5+3925 & 1.8 & -0.85  &  29.1 & 0.440$\pm$0.179 &  0.088   &  0.814$\pm$0.080 &  0.432\\
Mrk 266 SW    & 2.0 & -1.04  &  56.8 & 1.217$\pm$0.350 &  0.187   &  2.730$\pm$0.163 &  0.760\\
NGC 5273      & 1.9 & -0.22  &  18.5 & 0.167$\pm$0.219 &  0.017   &  0.707$\pm$0.114 &  0.511\\
IC 4329A      & 1.0 & -0.87  & 414.6 & 0.486$\pm$0.483 &  0.011   &  8.874$\pm$0.255 &  0.313\\
Mrk 279       & 1.0 & -0.80  & 103.4 & 0.187$\pm$0.181 &  0.012   &  0.051$\pm$0.077 &  0.006\\
Mrk 673       & 2.0 & -0.72  &  32.2 & 0.842$\pm$0.225 &  0.128   &  1.723$\pm$0.110 &  0.811\\
IC 4397       & 2.0 & -0.86  &  16.4 & 0.508$\pm$0.172 &  0.204   &  1.830$\pm$0.079 &  1.702\\
NGC 5548      & 1.5 & -0.81  & 175.5 & 0.762$\pm$0.355 &  0.037   &  0.348$\pm$0.170 &  0.026\\
NGC 5674      & 1.9 & -1.25  &  13.1 & 0.507$\pm$0.228 &  0.205   &  1.284$\pm$0.132 &  1.469\\
Mrk 817       & 1.5 & -0.85  & 121.8 &     $\leq$0.377 &$\leq$0.01&  1.452$\pm$0.160 &  0.171\\
NGC 5728      & 2.0 & -1.10  &  46.2 & 1.506$\pm$0.337 &  0.235   &  4.157$\pm$0.158 &  1.399\\
Mrk 841       & 1.0 & -0.99  &  66.9 & 0.374$\pm$0.221 &  0.041   &  0.698$\pm$0.120 &  0.137\\
NGC 5929      & 2.0 &  0.13  &  17.8 & 0.930$\pm$0.345 &  0.155   &  0.783$\pm$0.163 &  0.524\\
NGC 5940      & 1.0 & -1.40  &  27.2 &     $\leq$0.229 &$\leq$0.09&  1.399$\pm$0.115 &  0.796\\
NGC 5953      & 2.0 & -0.56  &  98.0 & 1.416$\pm$0.393 &  0.087   &  8.250$\pm$0.180 &  1.201\\
ESO 137-G34   & 2.0 & -0.75  &  11.0 &     $\leq$1.040 &  0.880   &  0.717$\pm$0.300 &  0.880\\
Mrk 507       & 2.0 & -0.93  &  15.3 & 0.204$\pm$0.159 &  0.087   &  0.416$\pm$0.070 &  0.409\\
H 1821+643    & QSO & -0.99  & 122.5 & 0.567$\pm$0.355 &  0.033   &  1.012$\pm$0.161 &  0.121\\
\\
\hline
\end{tabular}
\end{flushleft}
\label{flux1}
\end{table*}
\addtocounter{table}{-1}
\begin{table*}
 \caption[]{Continuum and PAH band fluxes (cont.)} 
  \begin{flushleft}
\begin{tabular}{lccccccc}
\hline
\hline
\\
TARGET        & Type&$\alpha$&F(7$\mu$m)&  F(3.3$\mu$m)  &EW(3.3$\mu$m)&  F(6.2$\mu$m)   &EW(6.2$\mu$m)\\
              &     &        & 
		(mJy) & 
	      ${\rm (10^{-12}\frac{erg}{cm^{2}\,s})}$ & 
		($\mu$m) & 
	      ${\rm (10^{-12}\frac{erg}{cm^{2}\,s})}$ & 
		($\mu$m) \\
\\
\hline
\\	       
3C 382        & 1.0 & -0.38  &  52.7 & 0.396$\pm$0.149 &  0.033   &  0.584$\pm$0.080 &  0.156\\
3C 390.3      & 1.0 & -0.53  &  52.1 &     $\leq$0.167 &$\leq$0.01&  0.048$\pm$0.080 &  0.027\\
ESO 141-G55   & 1.0 & -0.97  & 108.8 & 0.014$\pm$0.229 &  0.014   &  0.720$\pm$0.113 &  0.091\\
Mrk 509       & 1.0 & -0.84  & 161.2 &     $\leq$0.166 &$\leq$0.01&  0.624$\pm$0.079 &  0.055\\
NGC 7314      & 1.9 & -1.05  &  43.1 & 0.905$\pm$0.321 &  0.219   &  0.677$\pm$0.151 &  0.205\\
IR 22377+0747 & 1.8 & -0.50  &  36.6 &     $\leq$0.307 &$\leq$0.07&  0.521$\pm$0.166 &  0.201\\
Ark 564       & 1.0 & -1.18  &  84.8 &     $\leq$0.157 &$\leq$0.01&  0.405$\pm$0.070 &  0.069\\
NGC 7592      & 2.0 & -0.73  &  72.6 & 0.131$\pm$0.245 &  0.034   &  4.127$\pm$0.112 &  0.872\\
NGC 7603      & 1.5 & -0.84  &  99.8 & 0.321$\pm$0.344 &  0.019   &  1.603$\pm$0.174 &  0.229\\
NGC 7674      & 2.0 & -0.96  & 186.9 & 1.401$\pm$0.469 &  0.053   &  3.336$\pm$0.226 &  0.279\\
\\
\hline
\end{tabular}
\end{flushleft}
\end{table*}

\begin{table*}
  \caption[]{PAH band fluxes}
  \begin{flushleft}
\begin{tabular}{lccccccc}
\hline
\hline
\\
TARGET        &Type&    F(7.7$\mu$m)  &EW(7.7$\mu$m)&   F(8.6$\mu$m)   &EW(8.6$\mu$m)&${\rm F(PAH)}$&${\rm EW(PAH)}$\\
              &    & 
	      ${\rm (10^{-12}\frac{erg}{cm^{2}\,s})}$ & 
	      ($\mu$m) & 
	      ${\rm (10^{-12}\frac{erg}{cm^{2}\,s})}$ & 
	      ($\mu$m) & 
	      ${\rm (10^{-12}\frac{erg}{cm^{2}\,s})}$ & 
	      ($\mu$m) \\
\\
\hline
\\             
Mrk 334       &1.8 &  9.091$\pm$0.201&   2.063    &  2.649$\pm$0.124&	0.615	 &  14.547$\pm$0.276  &   3.191\\
Mrk 335       &1.0 &  1.801$\pm$0.303&   0.225    &  0.701$\pm$0.188&	0.104	 &   3.694$\pm$0.419  &   0.417\\
Fairall 9     &1.0 &  3.217$\pm$0.270&   0.410    &  0.400$\pm$0.184&	0.045	 &   4.790$\pm$0.376  &   0.575\\
NGC 526a      &1.5 &  2.596$\pm$0.191&   0.507    &  0.890$\pm$0.112&	0.197	 &   6.025$\pm$0.257  &   1.050\\
NGC 701       & SB & 15.711$\pm$0.121&   5.876    &  4.291$\pm$0.079&	1.627	 &  24.095$\pm$0.166  &   8.940\\
Mrk 590       &1.5 &  3.710$\pm$0.249&   0.555    &  1.011$\pm$0.155&	0.153	 &   5.817$\pm$0.342  &   0.860\\
NGC 1097      &2.0 & 79.759$\pm$0.444&   3.612    & 19.387$\pm$0.277&	0.897	 & 118.534$\pm$0.627  &   5.292\\
NGC 1125      &2.0 &  5.000$\pm$0.172&   2.405    &  1.161$\pm$0.116&	0.620	 &   7.716$\pm$0.245  &   3.530\\
NGC 1241      &2.0 &  3.708$\pm$0.121&   2.084	  &  1.088$\pm$0.082&	0.612	 &   5.390$\pm$0.171  &   2.983\\
NGC 1386      &2.0 & 10.217$\pm$0.355&   0.986	  &  2.349$\pm$0.235&	0.208	 &  14.737$\pm$0.504  &   1.366\\
NGC 1566      &1.0 & 10.562$\pm$0.604&   1.695	  &  2.601$\pm$0.382&	0.425	 &  16.990$\pm$0.889  &   2.617\\
NGC 1667      &2.0 & 14.707$\pm$0.257&   4.683	  &  3.955$\pm$0.164&	1.334	 &  22.136$\pm$0.354  &   6.981\\
Ark 120       &1.0 &  2.634$\pm$0.177&   0.377	  &  0.567$\pm$0.124&	0.108	 &   4.564$\pm$0.252  &   0.622\\
IR 05189-2524 &2.0 & 13.817$\pm$0.182&   1.104	  &  3.845$\pm$0.117&	0.355	 &  21.807$\pm$0.250  &   1.706\\
MCG 8-11-11   &1.5 &  3.414$\pm$0.175&   0.289	  &  1.013$\pm$0.118&	0.094	 &   5.056$\pm$0.249  &   0.403\\
Mrk 3         &2.0 &  4.138$\pm$0.171&   0.545	  &  1.108$\pm$0.111&	0.139	 &   5.938$\pm$0.240  &   0.785\\
HS 0624+6907  &QSO &  0.879$\pm$0.269&   0.275	  &  0.000$\pm$1.000&	0.000	 &   0.000$\pm$1.000  &   0.000\\
NGC 3227      &1.5 & 16.041$\pm$0.369&   1.581	  &  4.451$\pm$0.229&	0.448	 &  24.880$\pm$0.506  &   2.392\\
A 1058+45     &2.0 &  2.901$\pm$0.127&   3.247	  &  0.932$\pm$0.081&	1.132	 &   4.684$\pm$0.175  &   5.009\\
NGC 3516      &1.5 &  2.364$\pm$0.368&   0.161	  &  0.561$\pm$0.235&	0.046	 &   4.220$\pm$0.520  &   0.261\\
NGC 3982      &2.0 & 10.941$\pm$0.378&   2.405	  &  2.865$\pm$0.216&	0.657	 &  16.295$\pm$0.513  &   3.509\\
NGC 4051      &1.0 &  6.209$\pm$0.352&   0.428	  &  1.870$\pm$0.222&	0.136	 &   9.345$\pm$0.502  &   0.648\\
Mrk 766       &1.5 &  2.912$\pm$0.382&   0.285	  &  0.630$\pm$0.224&	0.045	 &   3.864$\pm$0.516  &   0.342\\
NGC 4388      &2.0 & 16.496$\pm$0.363&   1.612	  &  5.024$\pm$0.252&	0.556	 &  25.759$\pm$0.522  &   2.405\\
NGC 4507      &2.0 &  3.729$\pm$0.358&   0.243	  &  1.082$\pm$0.191&	0.053	 &   6.596$\pm$0.484  &   0.401\\
NGC 4579      &1.9 &  4.738$\pm$0.363&   0.945	  &  0.957$\pm$0.156&	0.254	 &   7.277$\pm$0.432  &   1.466\\
NGC 4593      &1.0 &  3.145$\pm$0.254&   0.338	  &  0.898$\pm$0.163&	0.104	 &   4.895$\pm$0.358  &   0.458\\
IR 12495-1308 &1.0 &  1.691$\pm$0.195&   0.760	  &  0.745$\pm$0.120&	0.357	 &   3.036$\pm$0.274  &   1.383\\
NGC 5033      &1.9 & 16.933$\pm$0.364&   2.394	  &  3.910$\pm$0.220&	0.571	 &  25.163$\pm$0.492  &   3.454\\
MCG-6-30-15   &1.0 &  1.469$\pm$0.420&   0.120	  &  1.599$\pm$0.339&	0.157	 &   3.193$\pm$0.610  &   0.298\\
Z 1335.5+3925 &1.8 &  3.667$\pm$0.124&   2.413	  &  1.190$\pm$0.083&	0.835	 &   5.692$\pm$0.173  &   3.619\\
Mrk 266 SW    &2.0 & 11.185$\pm$0.274&   3.545	  &  2.724$\pm$0.172&	0.843	 &  16.777$\pm$0.370  &   5.267\\
NGC 5273      &1.9 &  0.930$\pm$0.171&   1.028	  &  0.454$\pm$0.111&	0.619	 &   2.090$\pm$0.239  &   2.161\\
IC 4329A      &1.0 & 17.157$\pm$0.397&   0.714	  &  4.371$\pm$0.242&	0.204	 &  31.753$\pm$0.546  &   1.313\\
Mrk 279       &1.0 &  0.758$\pm$0.118&   0.135	  &  0.175$\pm$0.084&	0.036	 &   0.942$\pm$0.168  &   0.169\\
Mrk 673       &2.0 &  7.961$\pm$0.183&   4.617	  &  2.077$\pm$0.117&	1.319	 &  11.944$\pm$0.251  &   6.843\\
IC 4397       &2.0 &  6.449$\pm$0.126&   7.199	  &  1.866$\pm$0.077&	2.275	 &  10.307$\pm$0.173  &  11.422\\
NGC 5548      &1.5 &  4.019$\pm$0.267&   0.425	  &  1.204$\pm$0.165&	0.099	 &   5.238$\pm$0.363  &   0.552\\
NGC 5674      &1.9 &  3.400$\pm$0.209&   4.654	  &  1.616$\pm$0.135&	2.307	 &   6.365$\pm$0.291  &   8.520\\
Mrk 817       &1.5 &  3.561$\pm$0.251&   0.542	  &  0.888$\pm$0.169&	0.146	 &   6.081$\pm$0.353  &   0.871\\
NGC 5728      &2.0 & 15.579$\pm$0.261&   6.064	  &  3.053$\pm$0.158&	1.276	 &  23.420$\pm$0.358  &   8.835\\
Mrk 841       &1.0 &  2.331$\pm$0.190&   0.627	  &  0.856$\pm$0.120&	0.262	 &   3.879$\pm$0.260  &   1.044\\
NGC 5929      &2.0 &  2.472$\pm$0.242&   2.855	  &  0.956$\pm$0.156&	1.385	 &   4.256$\pm$0.347  &   4.697\\
NGC 5940      &1.0 &  3.089$\pm$0.186&   2.008	  &  1.075$\pm$0.120&	0.674	 &   5.532$\pm$0.257  &   3.460\\
NGC 5953      &2.0 & 33.820$\pm$0.279&   6.442	  &  8.351$\pm$0.171&	1.631	 &  49.869$\pm$0.375  &   9.430\\
ESO 137-G34   &2.0 &  4.040$\pm$2.000&   6.350	  &  0.850$\pm$0.420&	1.580	 &   6.240$\pm$3.100  &   9.560\\
Mrk 507       &2.0 &  2.035$\pm$0.121&   2.417	  &  0.589$\pm$0.102&	0.745	 &   3.154$\pm$0.179  &   3.675\\
H 1821+643    &QSO &  2.059$\pm$0.333&   0.318	  & -0.334$\pm$0.253&  -0.070	 &   3.079$\pm$0.459  &   0.371\\
\\
\hline
\end{tabular}
\end{flushleft}
\label{flux2}
\end{table*}
\addtocounter{table}{-1}
\begin{table*}
  \caption[]{PAH band fluxes (cont.)}
  \begin{flushleft}
\begin{tabular}{lccccccc}
\hline
\hline
\\
TARGET        &Type&    F(7.7$\mu$m)  &EW(7.7$\mu$m)&   F(8.6$\mu$m)   &EW(8.6$\mu$m)&${\rm F(PAH)}$&${\rm EW(PAH)}$\\
              &    & 
	      ${\rm (10^{-12}\frac{erg}{cm^{2}\,s})}$ & 
	      ($\mu$m) & 
	      ${\rm (10^{-12}\frac{erg}{cm^{2}\,s})}$ & 
	      ($\mu$m) & 
	      ${\rm (10^{-12}\frac{erg}{cm^{2}\,s})}$ & 
	      ($\mu$m) \\
\\
\hline
\\             
3C 382        &1.0 &  0.869$\pm$0.136&   0.297	  &  0.118$\pm$0.109&	0.057	 &   1.710$\pm$0.196  &   0.539\\
3C 390.3      &1.0 &  0.661$\pm$0.141&   0.223	  &  0.513$\pm$0.109&	0.267	 &   1.317$\pm$0.202  &   0.558\\
ESO 141-G55   &1.0 &  1.635$\pm$0.190&   0.265	  &  1.302$\pm$0.131&	0.257	 &   3.534$\pm$0.265  &   0.618\\
Mrk 509       &1.0 &  1.809$\pm$0.127&   0.206	  &  0.535$\pm$0.079&	0.075	 &   3.018$\pm$0.174  &   0.341\\
NGC 7314      &1.9 &  2.551$\pm$0.245&   1.046	  &  0.580$\pm$0.168&	0.170	 &   3.817$\pm$0.337  &   1.498\\
IR 22377+0747 &1.8 &  1.647$\pm$0.249&   0.906	  &  0.626$\pm$0.167&	0.377	 &   2.718$\pm$0.352  &   1.456\\
Ark 564       &1.0 &  1.258$\pm$0.115&   0.260	  &  0.187$\pm$0.073&	0.037	 &   1.870$\pm$0.159  &   0.368\\
NGC 7592      &2.0 & 15.973$\pm$0.189&   4.108	  &  4.086$\pm$0.113&	1.111	 &  24.389$\pm$0.256  &   6.136\\
NGC 7603      &1.5 &  4.247$\pm$0.266&   0.788	  &  1.168$\pm$0.183&	0.241	 &   7.139$\pm$0.378  &   1.260\\
NGC 7674      &2.0 & 11.509$\pm$0.376&   1.099	  &  3.503$\pm$0.238&	0.390	 &  18.942$\pm$0.516  &   1.811\\
\\
\hline
\end{tabular}
\end{flushleft}
\end{table*}

\end{document}